\documentclass[12pt]{iopart}

\usepackage{amsfonts}
\expandafter\let\csname equation*\endcsname\relax
\expandafter\let\csname endequation*\endcsname\relax
\usepackage{amsmath}
\usepackage{amssymb}
\usepackage[square, numbers, sort&compress, nonamebreak]{natbib}
\newcommand{\acknowledgments}[1]{\ack{#1}} 
\usepackage{graphicx}
\usepackage{color}
\usepackage{url}
\usepackage[bookmarks, breaklinks]{hyperref}

\newcommand{\ket}[1]{| #1 \rangle}                   
\newcommand{\bra}[1]{\langle #1 |}                   
\newcommand{\defeq}[0]{\equiv}                       
\DeclareMathSymbol{\R}{\mathalpha}{AMSb}{"52}        
\DeclareMathSymbol{\Q}{\mathalpha}{AMSb}{"51}        
\DeclareMathSymbol{\N}{\mathalpha}{AMSb}{"4E}        

\begin{document}

\title{Shortcuts to adiabaticity for trapped ultracold gases}

\author{Jean-Fran\c{c}ois Schaff$^1$, Pablo Capuzzi$^2$, Guillaume Labeyrie$^1$ and Patrizia Vignolo$^1$}
\address{$^1$ Universit\'{e} de Nice-Sophia Antipolis, Institut non lin\'{e}aire de Nice, CNRS, 1361 route des Lucioles, F-06560 Valbonne, France}
\address{$^2$ Universidad de Buenos Aires, FCEN, Departamento de Fisica and Instituto de Fisica de Buenos Aires, CONICET, Ciudad Universitaria, Pab. I C1428EGA Buenos Aires, Argentina}
\ead{jean-francois.schaff@inln.cnrs.fr}

\begin{abstract}
  We study, experimentally and theoretically, the controlled transfer
  of harmonically trapped ultracold gases between different quantum
  states.  In particular we experimentally demonstrate a fast
  decompression and displacement of both a non-interacting gas and an 
  interacting Bose-Einstein condensate which are initially at equilibrium.
  The decompression parameters are engineered such that the final state
  is identical to that obtained after a perfectly adiabatic transformation despite the fact that
  the fast decompression is performed in the strongly non-adiabatic regime. During the transfer
  the atomic sample goes through strongly out-of-equilibrium states while the external confinement is
  modified until the system reaches the desired stationary state. The
  scheme is theoretically based on the invariants of motion and
  scaling equations techniques and can be generalized to decompression
  trajectories including an arbitrary deformation of the trap. It is also directly applicable
  to arbitrary initial non-equilibrium states.
\end{abstract}


\section*{Introduction}

In Quantum Mechanics, the evolution of a system described by a
time-dependent Hamiltonian $H(t)$ is \emph{adiabatic} when the
transition probabilities between the instantaneous eigenstates of $H$
are negligible. This happens when $H$ is either time-independent, or
when its rate of change is \emph{slow} compared to the typical
time-scales involved~\cite{Born1928, Kato1950, Comparat2009}. Nevertheless, thinking
in terms of instantaneous eigenstates is often much easier than
looking for the solutions of time-dependent problems. In the field of
atomic physics, going from the semi-classical
approach of atom-field interaction to the celebrated dressed state
picture~\cite{Cohen-Tannoudji1992} illustrates the convenience of such adiabatic representations.

For this reason, many adiabatic schemes to prepare interesting quantum
states were proposed. For instance, non-classical states~\cite{Parkins1993, Cirac1994}, or new
strongly correlated states~\cite{Sorensen2010} can be
prepared by adiabatic passage. Quantum adiabatic
computation has recently been demonstrated~\cite{Peng2008}. Yet
adiabatic techniques are typically slow~\cite{Comparat2009}, while
experimentalist are often constrained by finite lifetimes or coherence
times of their samples. This motivated the search for fast schemes
reproducing or approaching adiabatic transformations. Some methods use
minimization techniques to optimize the transition to a target
state~\cite{Hohenester2007, DeChiara2008, Vasilev2009, Mundt2009},
whereas others yield the exact same state that would have been reached
after an adiabatic transformation~\cite{Berry2009, Chen2010}. The latter are referred to as \emph{shortcuts to adiabaticity}. In this
article, we detail how such methods can be used on the motional
degrees of freedom of ultracold gases confined in time-dependent
harmonic traps, and experimentally demonstrate the validity of the
approach.
Two direct applications of the procedure are the fast cooling of atomic samples, and the suppression (or reduction) of any \textit{parasitic excitations} which occur in experiments on ultracold 
gases when the trap geometry or the interactions are modified. Since the
method is not restricted to equilibrium states it could be used in a 
variety of situations as discussed at the end of the paper.

The first part is theoretical and recalls how harmonically confined
gases react to the variation of the trap. Both the one-dimensional
non-interacting gas, and the three-dimensional Bose-Einstein
condensate with repulsive contact interaction between particles are
treated. In the second part, the method to realize shortcuts to
adiabaticity are detailed for theses two systems, and examples are
given. The third part focuses on the experimental realization of these
methods. Rapid decompressions have been performed on both a
non-interacting gas and a Bose-Einstein condensate. The practical
limitations which degrade the results are discussed. In the last part
of the article, we attempt to generalize the problem to an arbitrary
variation of the three-dimensional harmonic potential and give other
examples of shortcuts which may be of experimental relevance.

\section{Scaling properties of harmonically confined ultracold gases: two examples}
\label{sec:scaling}

In this section, we recall how the density and velocity distributions
of a one-dimensional (1D) non-interacting gas are affected by a change
of the harmonic confinement. In 1D, the harmonic trap is fully
described by its time-dependent angular frequency $\omega(t)$, and
minimum position $q_0(t)$. We show that the dynamics is fully
described by two scaling functions, one associated to the cloud's
size, the other to its centre-of-mass position, and exhibit the exact
solutions of the Schr\"{o}dinger equation. This will be used in the
rest of the paper to realize shortcuts to adiabaticity
(cf. Sec.~\ref{sec:concept_shortcut}). Similar scaling properties are
also recalled for Bose-Einstein condensates (BECs) with strong 
interactions in the Thomas-Fermi regime. The analogy between the invariant
method used for the non-interacting gas~\cite{Lewis1969}, and the scaling often used for
BECs~\cite{Shlyapnikov1996, Castin1996, Shlyapnikov1997} is underlined.

\subsection{The non-interacting gas}
\label{sec:ni_gas}

We consider a 1D non-interacting gas confined in the most general
time-dependent harmonic potential, described by the one particle
Hamiltonian
\begin{equation}
H(q,p,t) = \frac{p^2}{2m} + \frac{1}{2} m \, \omega^2(t) \left[q - q_0(t)\right]^2 ,
\label{eq:hamiltonian}
\end{equation}
where $q$ and $p$ are conjugate variables, and $m$ is the mass of a
particle.  We first recall how dynamical invariants can be used to
find the general solutions of the Schr\"{o}dinger equation.

\subsubsection{Definition and properties of dynamical invariants}
\label{sec:invariants}

In 1969 \citet{Lewis1969} generalized the concept of invariant of
motion to the case of explicitly time-dependent Hamiltonians
$H(q,p,t)$. Such Lewis invariants (also called dynamical invariants,
or first integrals) can be used to solve the Schr\"{o}dinger equation
\begin{equation}
\label{eq:schrodinger}
i \hbar \frac{\partial \ket{t}}{\partial t} = H(q,p,t) \ket{t} .
\end{equation}

Given a time-dependent Hamiltonian $H(q,p,t)$, a time-dependent
hermitian operator $I(q,p,t)$ is a dynamical invariant of the system
described by $H$ if it is constant under Hamiltonian evolution, that
is if
\begin{equation}
\label{eq:def_invariant}
\frac{d I}{d t} \defeq \frac{\partial I}{\partial t} + \frac{1}{i \hbar} [I,H] = 0 \, .
\end{equation}
In this case, the following properties hold \cite{Lewis1969}:
\begin{enumerate}
\item if $\ket{t}$ is a solution of \eqref{eq:schrodinger}, then
  $I\ket{t}$ is also a solution of \eqref{eq:schrodinger},
\item the eigenvalues $\lambda(t)$ and associated eigenvectors
  $\ket{\lambda \, ; \, t}$ of $I$ are \textit{a priori}
  time-dependent. We assume they form a complete set. It turns out that the
  eigenvalues are actually constant: $\lambda(t) = \lambda$. They are
  real because $I$ is hermitian.
\item The eigenvectors of $I$ satisfy
\begin{equation}
\label{eq:almost}
\text{for all} \ \lambda, \lambda' \ \text{such that} \ \lambda \neq
\lambda', \quad 
\bra{\lambda' \, ; \, t} i \hbar \frac{\partial}{\partial t} \ket{\lambda \, ; \, t} =  \bra{\lambda' \, ; \, t} H \ket{\lambda \, ; \, t} .
\end{equation}
\item If we assume that $I$ does not contain the operator
  $\partial/\partial t$, then the set of vectors $\{ e^{i
    \alpha_{\lambda}(t)} \ket{\lambda \, ; \, t}, \
  \alpha_{\lambda}(t) \in \R(t) \}$ is also a complete set of
  eigenvectors of $I$. If these functions are chosen to solve the
  equations
\begin{equation}
\label{eq:phases}
\frac{d \alpha_{\lambda}}{d t} = \bra{\lambda \, ; \, t} i \frac{\partial}{\partial t} - \frac{H}{\hbar} \ket{\lambda \, ; \, t}
\end{equation}
then Eq.~\eqref{eq:almost} also holds for $\lambda' = \lambda$. Using
the fact that the set is complete, this gives the general solutions of
the time-dependent Schr\"odinger equation as
\begin{equation}
\ket{t} = \sum_{\lambda} c_{\lambda} \, e^{i \alpha_{\lambda}(t)} \ket{\lambda \, ; \, t} \, ,
\end{equation}
where the $c_{\lambda}$'s are constant complex numbers.
\end{enumerate}
The solutions of the Schr\"{o}dinger equation are thus given by the
knowledge of an invariant $I(q,p,t)$, any set of its time-dependent
eigenvectors, and the phases $\alpha_{\lambda}(t)$ which must solve
Eqs.~\eqref{eq:phases}.

\subsubsection{Derivation of a dynamical invariant}

In this section, we give a simple derivation of the invariants of a 1D time-dependent
harmonic oscillator (HO) described by~\eqref{eq:hamiltonian}. We use
the classical formalism to derive the invariant, which is also an
invariant of the corresponding quantum system.

The canonical equations of motion associated with the Hamiltonian~\eqref{eq:hamiltonian} are
\begin{subequations}
\begin{align}
\frac{dq}{dt} &= \{q, H\} = \frac{p}{m} , \label{eq:q} \\ 
\frac{dp}{dt} &= \{p, H\} = - m \, \omega^2(t) [q - q_0(t)] , \label{eq:p}
\end{align}
\end{subequations}
where $\{A, B\} \defeq \dfrac{\partial A}{\partial q} \dfrac{\partial
  B}{\partial p} - \dfrac{\partial A}{\partial p} \dfrac{\partial
  B}{\partial q}$ are the Poisson brackets of two observables $A$ and
$B$.

When the angular frequency $\omega(t)$ and trap centre $q_0(t)$ vary,
one expects the cloud to be displaced and to change its size, thus one
can introduce a canonical change of variables
\begin{equation}
Q = \frac{q-q_{\text{cm}}(t)}{b(t)} , \quad P = P(q,p,t) , \quad \tau = \tau(t) ,
\label{eq:canonical}
\end{equation}
leading to a new Hamiltonian $H'$. One has to derive conditions on the
real dimensionless function $b$, and the displacement function
$q_{\text{cm}}$ such that the transformation is canonical. For this,
we look for a new Hamiltonian of the form
\begin{equation}
H' = \frac{P^2}{2m} + \frac{1}{2} m \, \omega_0^2 \, Q^2 + f(\tau),
\end{equation}
where $\omega_0$ is a constant angular frequency. The Hamiltonian
explicitly depends on time only through the function $f(\tau)$ (which
does not contain the variables $Q$ and $P$).  The transformation
\eqref{eq:canonical} is canonical if
\begin{subequations}
\begin{align}
\frac{d Q}{d\tau} &= \{Q, H'\} , \label{cond1} \\
\frac{d P}{d\tau} &= \{P, H'\} . \label{cond2}
\end{align}
\end{subequations}
From Eq.~\eqref{cond1} one deduces that
\begin{equation}
d\tau = b^{-2} \, dt
\label{eq:tau}
\end{equation}
and that
\begin{equation}
P = b(p - m \dot{q}_{\text{cm}}) - m \dot{b} (q - q_{\text{cm}}),
\end{equation}
where $\dot{}$ denotes the derivation with respect to time $t$.
From Eq.~\eqref{cond2}, one finds the functions $b$ and $q_{\text{cm}}$ must obey the two differential equations
\begin{gather}
\ddot{b} + \omega^2(t) \, b = \frac{\omega^2_0}{b^3} , \label{eq:b} \\
\ddot{q}_{\text{cm}} + \omega^2(t) [q_{\text{cm}}(t) - q_0(t)] = 0 \label{eq:qcm} .
\end{gather}
When these two equations are satisfied, the quantity
\begin{equation}
I = \frac{P^2}{2m} + \frac{1}{2} m \, \omega_0^2 \, Q^2
\label{eq:invariant}
\end{equation}
which appears in the new Hamiltonian is a Lewis invariant. This can be
proved directly by checking that Eq.~\eqref{eq:def_invariant} is
verified.

The choice of the function $f(\tau)$ in $H'$ is irrelevant for the dynamics, since doing the change of Hamiltonian
\begin{equation}
H' \rightarrow H' - f(\tau) = I
\end{equation}
corresponds to a gauge transformation which changes the phase of the wave function in the following manner:
\begin{equation}
\psi_{H'}(Q,\tau) \rightarrow \psi_{I}(Q,\tau) = e^{\frac{i}{\hbar} F(\tau)} \psi_{H'}(Q,\tau) ,
\end{equation}
where $F$ is a primitive of $f$.

\subsubsection{Wave functions}

Once an invariant has been found, the results of
section~\ref{sec:invariants} can be used to calculate the wave
functions of the time-dependent HO~\eqref{eq:hamiltonian}. We use
Dirac's method to calculate the wave function of the time-independent
HO~\eqref{eq:invariant}. We define dimensionless variables
\begin{equation}
\xi = \sqrt{\frac{m \, \omega_0}{\hbar}} Q , \quad \pi = \frac{1}{\sqrt{m\hbar\omega_0}} P ,
\label{eq:dimless_var}
\end{equation}
satisfying the commutation relation $[\xi, \pi] = i$, and introduce the lowering and raising operators
\begin{equation}
a = \frac{1}{\sqrt{2}}(\xi + i \pi), \quad a^{\dagger} = \frac{1}{\sqrt{2}}(\xi - i \pi) .
\end{equation}
The invariant reads
\begin{equation}
I = \hbar \omega_0 (a^{\dagger}a + 1/2) .
\end{equation}
The eigenstates $\ket{n}$ of the number operator $\hat{n} \defeq a^{\dagger}a$ are the eigenstates of $I$ and satisfy
\begin{equation}
\quad a \ket{n} = \sqrt{n} \ket{n-1}, \quad a^{\dagger} \ket{n} = \sqrt{n+1} \ket{n+1}, \quad n \in \N .
\end{equation}
The eigenvalues of $I$ are
\begin{equation}
\lambda_n = \left(n + \frac{1}{2} \right) \hbar \omega_0, \quad n \in \N .
\end{equation}
The wave function $\psi_0(q,t) \defeq \langle q | 0 \rangle$ is calculated by solving the equation
\begin{equation}
a\ket{0} = 0
\end{equation}
in $\ket{q}$ representation.  The expression of $\pi$ is obtained from
$p = - i \hbar \, \partial/\partial q$, and
Eqs.~\eqref{eq:dimless_var} and \eqref{eq:canonical}, and reads
\begin{equation}
  \pi = -i\frac{\partial}{\partial \xi} - \frac{b\dot{b}}{\omega_0} \xi - \sqrt{\frac{m}{\hbar \omega_0}} \, b \, \dot{q}_{\text{cm}} .
\end{equation}
Imposing the normalization condition $\int dq \, |\psi_0(q,t)|^2 = 1$,
and calculating the time-dependent phase corresponding to the initial
Hamiltonian~\eqref{eq:hamiltonian}, we obtain the wave function
\begin{equation}
  \psi_0(q,t) = \frac{\pi^{-1/4}}{\sqrt{a_\text{ho} b}} \exp \left[ -\frac{1}{2} \left(\frac{q-q_{\text{cm}}}{a_\text{ho} b}\right)^2 \right] 
  e^{- \frac{i}{\hbar} F(t)}  
  e^{ i \phi(q,t) } 
  e^{- \frac{i}{\hbar} \lambda_0 \int_0^t dt'/b^2}
\end{equation}
where
\begin{gather}
  \phi(q,t) = \frac{m}{\hbar} \left[ \frac{\dot{b}}{2b} q^2 + \frac{1}{b} \left(\dot{q}_{\text{cm}} b - q_{\text{cm}} \dot{b} \right) q \right] , \\
  F(t) = \frac{m}{2} \int_0^t dt' \left[ \frac{1}{b^2}
    \left(\dot{q}_{\text{cm}} b - q_{\text{cm}} \dot{b}\right)^2 -
    \omega_0^2 \frac{q_{\text{cm}}^2}{b^4} + \omega^2(t') q_0^2
  \right] ,
\end{gather}
and $q_0$, $b$, $q_{\text{cm}}$, and their derivatives are functions
of $t$ ($t'$ when they are under an integral symbol) and are linked by
Eqs.~\eqref{eq:b} and \eqref{eq:qcm}. $a_\text{ho} = \sqrt{\hbar/m \omega_0}$ is the HO
length of $I$.

From this expression, we see the physical meaning of the two scaling
functions: $q_\text{cm}(t)$ is the centre of the wave function
(centre of mass of a cloud which was initially at equilibrium), and
$a_\text{ho} b$ is the width of the wave function.

The wave function associated to the eigenvalue $\lambda_n$ of $I$ is
expressed in terms of the nth Hermite polynomial $H_n$ as
\begin{equation}
  \psi_n(q,t) = \frac{1}{2^n n!} \psi_0(q,t) H_n\left(\frac{q-q_{\text{cm}}}{a_\text{ho} b}\right) e^{- \frac{i}{\hbar} (\lambda_n - \lambda_0) \int_0^t dt'/b^2} .
\end{equation}

\subsection{The case of an interacting Bose-Einstein condensate}

For the corresponding three-dimensional (3D) interacting system of $N$
particles, the Hamiltonian is
\begin{equation}
H = \sum_{i=1}^N \left[ \frac{\mathbf{p}_i^2}{2 m} + U(\mathbf{r}_i,t) \right] + \sum_{i<j} V(\mathbf{r}_j - \mathbf{r}_i) .
\end{equation}
The potential $U$ is supposed to be a time-dependent 3D HO, and the
rotation of this harmonic confinement is excluded for the moment (the
trap keeps the same eigenaxes):
\begin{equation}
U(\mathbf{r},t) = \frac{1}{2} m \Big\{ \omega_x^2(t) \left[ r_x - r_x^0(t) \right]^2
+ \omega_y^2(t) \left[ r_y - r_y^0(t) \right]^2 + \omega_z^2(t) \left[ r_z - r_z^0(t) \right]^2 \Big\} ,
\label{eq:pot}
\end{equation}
$V$ is the interaction potential between two particles, which is well approximated by a delta function for ultracold gases~\cite{Pethick2002}.

The procedure described in Sec.~\ref{sec:ni_gas} cannot be easily
adapted, because it would require the knowledge of an invariant of
this many-body system. But, when dealing with a BEC, the dynamics is
well described by a single particles wave function, whose evolution
obeys a non-linear Schr\"{o}dinger equation, the Gross-Pitaevskii
equation (GPE)~\cite{Pethick2002}.

\subsubsection{Scaling approach}
\label{sec:scaling-BEC}

Let us consider a quantum system described by the wave function
$\psi(\mathbf{r}, t)$, whose time evolution is given by the GPE
\begin{equation}
i \hbar \frac{\partial}{\partial t} \psi(\mathbf{r}, t) = \left[-\frac{\hbar^2}{2m} \Delta + U(\mathbf{r},t) + 
\tilde{V} N |\psi(\mathbf{r}, t)|^2 \right] \psi(\mathbf r, t) , \label{eq:gpe}
\end{equation}
with $m$, the mass of the particles, $N$ the number of particles, and
$\tilde{V} = 4 \pi \hbar^2 a_s/m$ the interaction coupling constant
generated by $s$-wave scattering between particles, characterized
by the scattering length $a_s$.  Analogously to the non-interacting
case, we wish to write the solution of the time-dependent GPE as
a function of the solution of a time-independent one expressed in a
suitable frame of reference. Following this line, a strategy to solve
Eq.~\eqref{eq:gpe} is to find a change of variables
$\boldsymbol{\rho}(\mathbf{r}, \{b_i(t)\}, \{r_i^{\text{cm}}(t)\})$
where the $b_i$'s and the $r_i^{\text{cm}}$'s are scaling and translation
functions such that Eq.~\eqref{eq:gpe} can be written as a
time-independent equation (i.e. a GPE with a \emph{time-independent}
potential) on the wave function $\chi(\boldsymbol{\rho},\tau)$,
defined by the relation
\begin{equation}
\psi(\mathbf r, t) = \mathcal{A}(t) \chi(\boldsymbol{\rho},\tau) e^{i\phi(\mathbf r, t)} ,
\label{defchi}
\end{equation}
$\mathcal{A}(t)$ being a time-dependent normalization factor and
$\phi(\mathbf r, t)$ a space- and time-dependent phase. All the
dynamics induced by the time-dependent potential is
\textit{transferred} to the functions $\{b_i(t)\}$ and
$\{r_i^{\text{cm}}(t)\}$, and the differential equations they have to
satisfy (similar to Eqs.~\eqref{eq:b} and ~\eqref{eq:qcm}).  If one
can solve the new time-independent equation on $\chi$, one solves
Eq.~\eqref{eq:gpe} and knows the wave function $\psi(\mathbf{r}, t)$.

Equation~\eqref{eq:gpe} is invariant under the transformation
\begin{equation}
\forall i \in \{x, y, z\}, \quad \rho_i = \frac{r_i - r_i^\text{cm}(t)}{b_i(t)}
\label{eq:change_rho}
\end{equation}
in any of the following cases:
\begin{enumerate}
\item in the non-interacting limit~\cite{Shlyapnikov1996,Muga2009}: in
  this case the system is equivalent to three independent HO of the
  kind treated in Sec.~\ref{sec:ni_gas},
\item for a suitable driving of the interaction term $\tilde{V}$
  \cite{Muga2009}, that is, assuming one can control $\tilde{V}(t)$ at
  will (for cold gases, this can be done using Feshbach resonances),
\item in the TF limit~\cite{Castin1996}.
\end{enumerate}
This third case, which is detailed in the following section, is
used in the rest of the paper.

\subsubsection{Condensate wave function in the Thomas-Fermi approximation}

Given a time-dependent Gross-Pitaevskii equation, the TF approximation
consists in neglecting the kinetic-energy-like term in the
$\boldsymbol{\rho}$-frame of reference, i.e. $-\hbar^2/(2m) \, \sum_i
b_i^{-2}\partial^2\chi/\partial \rho_i^2$, supposed to be small
compared to the interaction term~\cite{Castin1996,Shlyapnikov1997}.
In this regime, provided that $\mathcal{A}(t)=(\Pi_i b_i)^{-1/2}$ and
that
\begin{equation}
\phi(\mathbf{r},t)=\dfrac{m}{\hbar}
\left\{\sum_i\left[\dfrac{r_i^2}{2}\dfrac{\dot b_i}{b_i}+\dfrac{r_i}{b_i}\left(\dot r_i^{\text{cm}}b_i-r_i^{\text{cm}}\dot b_i\right)\right]\right\}+\phi_0(t),
\label{TF}
\end{equation}
with
\begin{equation}
\phi_0(t) = -\frac{m}{2\hbar}\sum_i\int_0^t{\rm d}t' \left\{
\frac{1}{b_i^2}\left(\dot{r}_i^{\text{cm}}b_i-r_i^{\text{cm}}\dot{b}_i
  \right)^2 \!\!- \omega_i^2(0)\frac{(r_i^{\text{cm}})^2}{b_i^4}+\left[\omega_i(t')r_i^0(t')\right]^2\right\}
\end{equation}
where the scaling and translation functions are solutions of
\begin{gather}
\forall i \in \{x, y, z\}, \quad \ddot{b}_i + \omega_i^2(t) b_i = \frac{\omega_i^2(0)}{b_i b_x b_y b_z} , \label{eq:size}\\
\ddot{r}_i^{\text{cm}} + \omega_i^2(t) \left[r_i^{\text{cm}}-r_i^0(t)\right] = 0 , \label{eq:position}
\end{gather}
one gets the following equation on $\chi$:
\begin{equation}
i\hbar\dfrac{\partial }{\partial\tau}\chi(\boldsymbol{\rho},\tau)=\left[ U(\boldsymbol{\rho}, 0) + \tilde{V} N |\chi(\boldsymbol{\rho},\tau)|^2 \right] \chi(\boldsymbol{\rho},\tau), \label{eq:gpe2}
\end{equation}
where we defined a rescaled time
\begin{equation}
\tau(t) = \int_0^t \frac{dt'}{\Pi_i b_i(t')} .
\label{eq:tau_bec}
\end{equation}
If at $t=0$ the
condensate is at equilibrium, the solution of Eq.~\eqref{eq:gpe2} is
\begin{equation}
\chi(\boldsymbol{\rho},\tau)=\left[\dfrac{\mu-U(\boldsymbol{\rho}, 0)}{\tilde{V} N }\right]^{1/2}e^{-i\frac{\mu}{\hbar}\tau},
\end{equation}
$\mu$ being the chemical potential. This gives the typical inverted
parabola density profile whose sizes evolve in time as
$R_i(t)=R_{i}(0) b_i(t)$, $R_{i}(0)=\sqrt{2\mu_0/m\omega_{i}^2(0)}$
being the initial TF radii.

\section{Shortcuts to adiabaticity}
\label{sec:concept_shortcut}

In this section the definition of a shortcut to adiabaticity is given,
and the results of Sec.~\ref{sec:scaling} are used to derive angular
frequency trajectories realizing such shortcuts, for both
non-interacting gases and interacting BECs confined in time-dependent
harmonic traps.

\subsection{Shortcut to adiabaticity based on an invariant of motion}

For a system described by a Hamiltonian $H(t)$, a \emph{shortcut to adiabaticity}
is realized when another Hamiltonian $H'(t)$ can be
found, such that the state obtained after a finite time of evolution
with $H'(t)$ is \emph{identical} (up to a global phase factor) to the
final state of the adiabatic evolution with $H(t)$. Shortcuts to
adiabaticity are \emph{not} adiabatic; only the final state is
identical to that obtained after an adiabatic evolution.

The possibility of such shortcuts has been known for a long time. For
instance, in the case of a HO with a time-dependent frequency
$\omega(t)$ treated in Ref.~\cite{Lewis1969}, when discussing the
transition probability $P_{sm}$ between two instantaneous eigenstates
$\ket{s \, ; \, t}$ and $\ket{m \, ; \, t}$, the authors noticed that
some trajectories $\omega(t)$ could lead to the same result as the
adiabatic case, namely
\begin{equation}
P_{sm} = \delta_{sm} .
\end{equation}
Such shortcuts to adiabaticity can thus be realized simply by
engineering the time-dependent parameters of the Hamiltonian.

A practical method to find a class of appropriate $\omega(t)$ was
detailed by \citet{Chen2010}. In this case, the Hamiltonian is chosen
to be time-independent (but with different frequencies) outside the time
interval $t \in [0, t_f]$. An invariant is engineered to commute with
the Hamiltonian outside this interval. This yields a specific
$\omega(t)$ for which all the eigenstates of $H(t<0)$ are exactly
mapped to the corresponding ones of $H(t>t_f)$ after the evolution
for $t \in [0, t_f]$. Up to a global phase and a rescaling of energies
and lengths, the final state (at time $t=t_f$) is identical to the
initial one ($t = 0$), i.e. if the initial state was
\begin{equation}
\ket{\psi \, ; \, t \leq 0} = \sum_n c_n \ket{n \, ; \, t=0} e^{- i \omega_n(0) t},
\end{equation}
where $\{\ket{n \, ; \, t}, n \in \N\}$ is a basis of instantaneous
eigenstates of $H(t)$ and $\{\hbar \omega_n(t)\}$ the corresponding eigenvalues, and $\sum_n |c_n|^2 = 1$, then the final state is
\begin{equation}
\ket{\psi \, ; \, t \geq t_f} = e^{i \Phi} \sum_n c_n \ket{n \, ; \, t_f} e^{- i \omega_n(t_f) t}.
\end{equation}
This is true even if the initial state is not an equilibrium state.

\subsubsection{Frequency trajectory for a non-interacting gas}
\label{sec:non-interacting_HO}

\begin{figure}[ht]
\begin{center}
\includegraphics[width=0.7\linewidth]{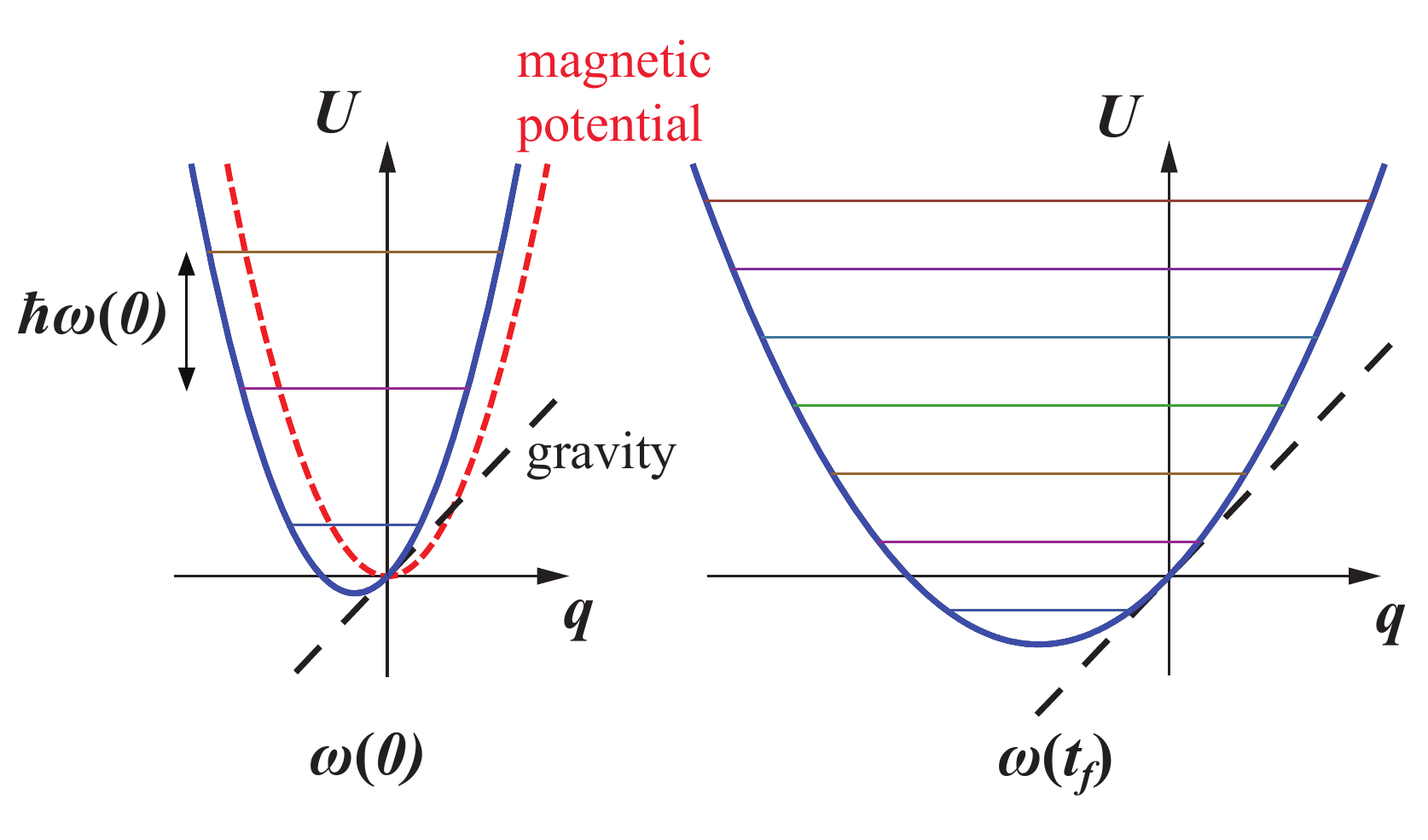}
\end{center}
\caption[Schematic representation of the trap
decompression.]{Schematic representation of the trap
  decompression. The potential (plain blue line) is the sum of the
  gravitational potential (dashed black line) and the harmonic
  magnetic potential (dashed red line). When the trap frequency is
  changed from $\omega(0)$ to $\omega(t_f)$, the lengths are
  multiplied by $\gamma = \sqrt{\omega(0)/\omega(t_f)}$, and the
  energies divided by $\gamma^2$. Because of gravity, the trap centre
  shifts vertically by $\Delta q = -g\left[1/\omega^2(t_f) -
  1/\omega^2(0)\right]$.}
\label{fig:scaling}
\end{figure}

The Hamiltonian is assumed to have the form 
\begin{equation}
H = \frac{p^2}{2m} + \frac{1}{2} m \, \omega^2(t) q^2 + mg q ,
\end{equation}
which is identical to~\eqref{eq:hamiltonian}, with the additional
constraint $q_0(t) = -g/\omega^2(t)$ (and a gauge transformation
consisting in adding $-m \, \omega^2(t) \, q_0^2(t)/2$ to $H$). It
describes a single particle in a harmonic trap subject to a constant
force, which, in the experiments presented in
Sec.~\ref{sec:experiments}, comes from gravity. The angular frequency
$\omega(t)$ is assumed to be constant outside the interval $t \in [0,
t_f]$. During this interval, the problem is to find the appropriate
frequency trajectory $\omega(t)$, connecting the initial trap of
initial frequency $\omega(0)$ to a final trap of frequency
$\omega(t_f)$, for the decompression (or compression if
$\omega(0)<\omega(t_f)$) to implement a shortcut to adiabaticity. Figure~\ref{fig:scaling} shows the initial and final situations assuming a decompression ($\omega(t_f)<\omega(0)$).

We used the strategy introduced by~\citet{Chen2010}. If the invariant commutes with the Hamiltonian
\begin{equation}
[I, H] = 0
\label{eq:commutate}
\end{equation}
for $t \leq 0$ and $t \geq t_f$, and provided that the functions $b$
and $q_\text{cm}$ are sufficiently continuous, the stationary states
of $H(t \leq 0)$ will be transferred to the corresponding ones of $H(t
\geq t_f)$.

It is convenient to use the dimensionless function
\begin{equation}
c(t) = -\frac{\omega_0^2}{g} \frac{q_\text{cm}(t)}{b(t)} 
\end{equation}
instead of $q_\text{cm}$, and to rewrite Eq.~\eqref{eq:qcm} using the rescaled time $\tau$ (Eq.~\eqref{eq:tau}). Equation~\eqref{eq:qcm} becomes
\begin{equation}
d^2 c / d\tau^2 + \omega_0^2 \, c = \omega_0^2 \, b^3 \label{eq:c(tau)} .
\end{equation}
If one chooses to set $\omega_0 = \omega(0)$, and the conditions
\begin{subequations} \label{eq:BCs1}
\begin{align}
b(0) &= 1, & \dot{b}(0) &= 0, \\
c(0) &= 1, & \dot{c}(0) &= 0,
\end{align}
then $I(0) = H(t \leq 0)$, and if
\begin{align}
b(t_f) &= \gamma,   & \dot{b}(t_f) &= 0, \\
c(t_f) &= \gamma^3, & \dot{c}(t_f) &= 0,
\end{align}
where $\gamma \defeq \sqrt{\omega_0/\omega_f}$, then $I(t_f) =
\gamma^2 H(t \geq t_f) + h(t)$, where $h$ is a function of time
only. These boundary conditions thus fulfil the
condition~\eqref{eq:commutate}. Since the functions $b$ and $c$ must
be solutions of Eqs.~\eqref{eq:b} and \eqref{eq:c(tau)}, four
additional boundary conditions must be satisfied:
\begin{align}
\ddot{b}(0) &= 0, & \ddot{b}(t_f) &= 0, \\
\ddot{c}(0) &= 0, & \ddot{c}(t_f) &= 0.
\end{align}
\end{subequations}
In order to construct the functions $b$ and $c$ satisfying these
boundary conditions and the two differential equations~\eqref{eq:b}
and \eqref{eq:c(tau)}, it is convenient to write all the boundary
conditions on the function $c$ and its derivatives with respect to the
rescaled time $\tau$. Using Eqs.~\eqref{eq:tau} and \eqref{eq:b}, and
differentiating Eq.~\eqref{eq:c(tau)} twice with respect to $\tau$,
one finds the ten conditions
\begin{subequations} \label{eq:BCs}
\begin{align}
c(0) &= 1, \label{eq:BC1}\\ c(\tau_f) &= \gamma^3, \label{eq:BC2}\\
\intertext{and, for all $k \in \{1,2,3,4\}$,}
\frac{d^{k}c}{d\tau^k}(0) &= 0, \label{eq:BC3}\\ \frac{d^{k}c}{d\tau^k}(\tau_f) &= 0 ,\label{eq:BC4}
\end{align}
which are sufficient for the twelve boundary
conditions~\eqref{eq:BCs1}. $\tau_f$ is the rescaled time
corresponding to $t_f$: $\tau_f = \int_0^{t_f} b^{-2}(t') \, dt'$.

Under this form, the boundary conditions are well suited to use a
polynomial ansatz for $c(\tau)$, deduce $b(\tau)$ with
Eq.~\eqref{eq:c(tau)}, compute $\tau(t)$ by numerically integrating
Eq.~\eqref{eq:tau}, and obtain $b(t)$. The final step consists in
using Eq.~\eqref{eq:b} to obtain the time-dependent trap frequency as
$\omega^2(t) = \omega^2_0/b^4 - \ddot{b}/b$.

An example of this procedure is given on Fig.~\ref{fig:c_b_t_nu} for
particular values of the initial and final frequencies. The final
rescaled time $\tau_f$ can be chosen at will, it can be arbitrarily
small, but one important constraint on the function $c$ is that it
must not lead to vanishing values of $b$ which give infinite
$\omega^2(t)$. Additional constraints on $c$ arise from experimental
requirements, such as positive $\omega^2(t)$ (attractive potentials),
maximal and minimal frequencies attainable with a given setup, \textit{speed}
at which $\omega(t)$ can be varied etc. Since all this depends on a
particular experimental setup, no mathematical analysis of the
\textit{best} ansatz to use was done.
\end{subequations}

For the experiments presented in Sec.~\ref{sec:experiments} and in
Refs.~\cite{Schaff2010b, Schaff2010c}, a polynomial of order fifteen
was used:
\begin{equation}
c(\tau) = \sum_{k=0}^{15} c_k \left( \frac{\tau}{\tau_f} \right)^k .
\label{eq:ansatz}
\end{equation}
The first coefficient is fixed to 1 by Eq.~\eqref{eq:BC1} and $c_1,
\cdots, c_4$ are fixed to 0 by Eqs.~\eqref{eq:BC3}. We arbitrarily
impose $c_5 = c_6 = \cdots = c_{10} = 0$, which leaves five coefficients
which are uniquely determined by the remaining boundary conditions
\eqref{eq:BC2} and \eqref{eq:BC4}. The calculation of these remaining
coefficient is done by inverting the linear system corresponding to
these five equations.

In principle, since there are ten BCs, a 9th order polynomial can be
used, which yields a unique solution for the ten coefficients of
$c$. Nevertheless, the obtained trajectory was not well behaved enough to be
realized experimentally (the frequency was decreasing too fast in the
beginning compared to what could be achieved by the apparatus). This is the
reason why a higher order polynomial was used and six coefficients were
fixed to 0.

Since the polynomial can be of any order greater than 9, and the boundary conditions only impose a linear relation between nine of its coefficients, there is obviously an infinity of different solutions connecting two given initial and final states. Moreover, other functions than polynomials could be used for $c$, as long as they provide enough free parameters.

The obtained nonzero coefficients of~\eqref{eq:ansatz} are given in table~\ref{tab:coefficients}.
\begin{table}[ht]
\caption[Nonzero coefficients of the fifteenth order polynomial ansatz.]{\label{tab:coefficients} Nonzero coefficients of the polynomial ansatz for $c(\tau)$ calculated from the boundary conditions~\eqref{eq:BCs}.}
\begin{indented}
\item[]\begin{tabular}{c|c|c|c|c|c}
$c_{0}$ & $c_{11}$           & $c_{12}$           & $c_{13}$           & $c_{14}$           & $c_{15}$           \\
\hline
1       & $1365(\gamma^3-1)$ & $5005(\gamma^3-1)$ & $6930(\gamma^3-1)$ & $4290(\gamma^3-1)$ & $1001(\gamma^3-1)$ \\
\end{tabular}
\end{indented}
\end{table}

\subsubsection{Example}
\label{sec:example_thermal}

\begin{figure}[ht]
\includegraphics[width=\linewidth]{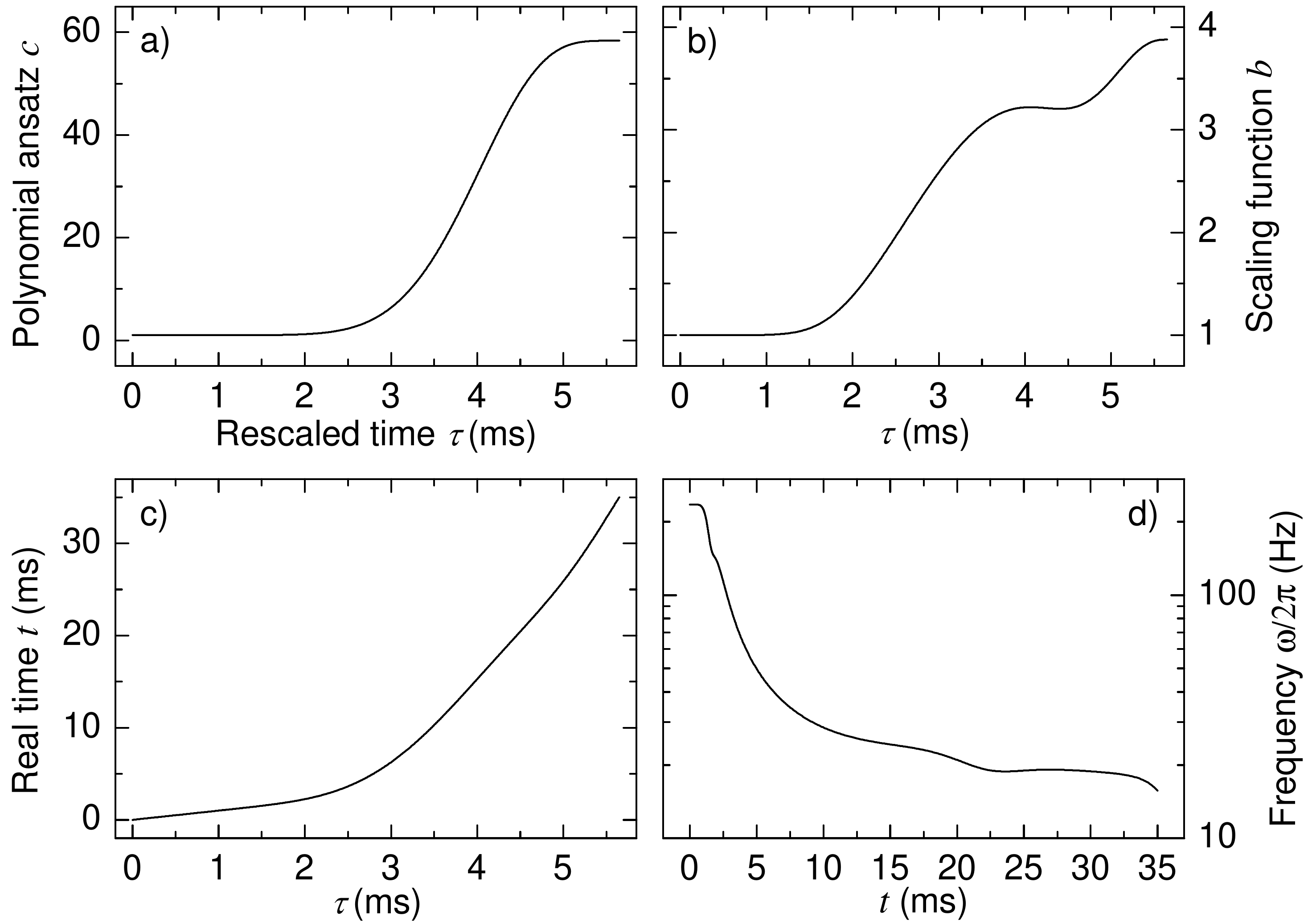}
\caption[Calculation of an optimal trap frequency
trajectory.]{Determination of the frequency trajectory when the trap
  is decompressed from $\omega(t=0)/2\pi = 235.8$~Hz to
  $\omega(t_f)/2\pi = 15.67$~Hz within $35$~ms (cf. parameters of
  Tab.~\ref{tab:param_th}). (a) A fifteenth order polynomial ansatz is
  used for the displacement function $c(\tau)$, which gives (b) the
  scaling function $b(\tau)$ through Eq.~\eqref{eq:c(tau)}; (c) real
  time $t(\tau)$ is calculated by numerically integrating
  Eq.~\eqref{eq:tau}; (d) Eq.~\eqref{eq:b} is used to determine the
  time-dependent frequency $\omega(t)/2\pi$ (notice the logarithmic
  scale).}
\label{fig:c_b_t_nu}
\end{figure}

In this section we determine the trajectory used in
Sec.~\ref{sec:experiment_th} and in Ref.~\cite{Schaff2010b}. The
parameters are given in
table~\ref{tab:param_th}. Figure~\ref{fig:c_b_t_nu} shows the
functions $c(\tau)$, $b(\tau)$, $t(\tau)$ and $\omega(t)/2\pi$
corresponding to this decompression.

\begin{table}[ht]
\caption[Decompression parameters for the non-interacting gas.]{\label{tab:param_th}Parameters of the 1D decompression of a non-interacting thermal gas.}
\begin{indented}
\item[]\begin{tabular}{r|l}
Initial frequency $\omega(0)/2\pi$ & $235.8$ Hz \\
Final frequency $\omega(t_f)/2\pi$ & $15.67$ Hz \\
Final rescaled time $\tau_f$       & $5.65$ ms \\
Corresponding duration $t_f$       & $35.0$ ms
\end{tabular}
\end{indented}
\end{table}

\begin{figure}[ht]
\includegraphics[width=\linewidth]{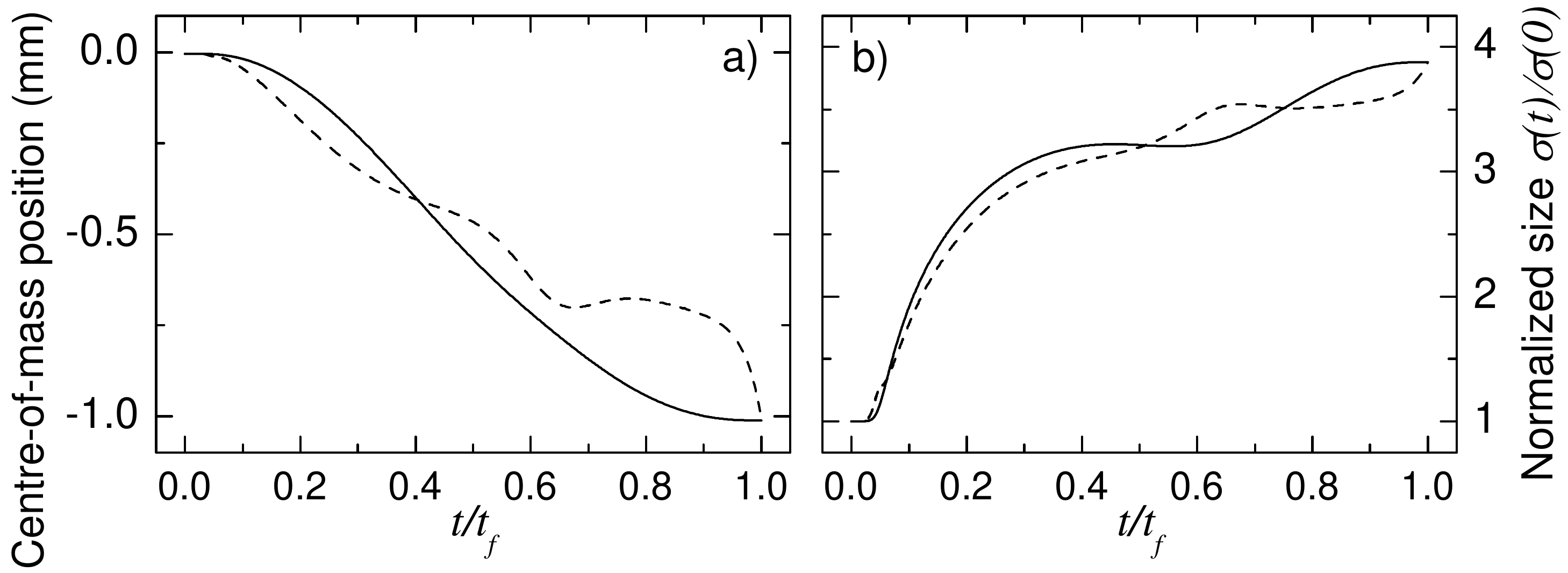}
\caption[Theoretical cloud size and centre-of-mass position during
decompression.]{Expected (a) centre-of-mass
  position and (b) cloud size during a fast decompression (same parameters as
  Tab.~\ref{tab:param_th} and Fig.~\ref{fig:c_b_t_nu}). The dashed
  curves correspond to the same values in the adiabatic limit $t_f
  \rightarrow \infty$. The adiabatic centre-of-mass position is the
  trap minimum $q_{\text{ad.}}(t) = -g / \omega^2(t)$, and the
  adiabatic size is $\sigma_{\text{ad.}}(t) =
  \sqrt{\omega_0/\omega(t)}\sigma(0)$.}
\label{fig:th_position_size}
\end{figure}

Since the exact wave functions are known, all the properties of the
atomic cloud can be calculated during decompression. For instance,
Fig.~\ref{fig:th_position_size} displays the size and centre-of-mass
position of a cloud initially at equilibrium in the compressed
trap. These are compared to the same values if the decompression were
done very slowly as in the adiabatic theorem. The clear difference between the plain and dashed curves illustrates the
fact that the decompression is not adiabatic.

\subsection{Shortcut to adiabaticity for an interacting Bose-Einstein condensate in the Thomas-Fermi limit}
\label{sec:interacting_bec}

Let us suppose that $\psi(\mathbf r, t \leq 0)$ is a stationary state of
Eq.~\eqref{eq:gpe}. We can engineer the parameters of the potential
$U(\mathbf r, t)$ such that $\psi(\mathbf r, t_f)$ is also a
stationary state for $t \geq t_f$. This implies that $\chi(\mathbf \rho,
\tau \geq \tau_f)$, with $\tau_f = \tau(t_f)$, must be a stationary state
of Eq.~\eqref{eq:gpe2} and that $\nabla_{\mathbf r} \phi(\mathbf r,
t_f)=0$. If these two conditions hold, $\psi(\mathbf r, t)$ can evolve
during the time interval $[0,t_f]$ between two stationary states even
being strongly different from the \textit{adiabatic stationary state}
during the evolution for $0 < t < t_f$.  In our experiment, the
time-dependent trapping potential has a cylindrical symmetry of the
form
\begin{equation}
  U(\mathbf r,t) = \frac{1}{2} m \, \omega_\perp^2(t)(r_x^2+r_z^2) + \frac{1}{2} m \, \omega_\parallel^2(t)r_y^2 + mg r_z\,,
  \label{trap}
\end{equation}
with initial and final angular frequencies
$\omega_{\perp,\parallel}(0)$ and $\omega_{\perp,\parallel}(t_f) =
\omega_{\perp,\parallel}(0)/\gamma^2_{\perp,\parallel}$, respectively.
This case corresponds to fix $\forall \ t, \ r_{x,y}^0(t)=0$ in Eq.~\eqref{eq:pot} and
$r_z^0(t)=-g/\omega_\perp^2(t)$.
By introducing the dimensionless function 
\begin{equation}
c(t) = -\frac{\omega_{\perp}^2(0)}{g} \frac{r_z^\text{cm}(t)}{b_\perp(t)} 
\label{ciditi}
\end{equation}
the differential equations \eqref{eq:size} and \eqref{eq:position} take the form
\begin{gather}
\label{eq:bperp}
\ddot b_\perp(t)+ b_\perp(t)\omega_\perp^2(t) = 
\omega_{\perp}^2(0)/[b_\perp^3(t)b_\parallel(t)] \\
\label{eq:bparallel}
\ddot b_\parallel(t)+ b_\parallel(t)\omega_\parallel^2(t) = 
\omega_{\parallel}^2(0)/[b_\parallel^2(t)b_\perp^2(t)]\\
\label{eq:a}
b_\perp^4(t)b_\parallel(t)\ddot c(t) + 2 b_\perp^3(t)
b_\parallel(t)\dot b_\perp(t)\dot c(t)
+\omega_{\perp}^2(0) c(t)-\omega_{\perp}^2(0) b_\perp^3(t)b_\parallel(t) = 0.
\end{gather}

The final state is an equilibrium state if the final TF radii verify
that $R_{\perp,\parallel}(t_f)/R_{\perp,\parallel}(0) =
\gamma_{\perp,\parallel}^2$, if the vertical centre-of-mass position
fulfils the condition $r_z^{\text{cm}}(t_f)/r_z^{\text{cm}}(0) = \gamma_\perp^4$, and if
the condensate flow is null, namely if ${\mathbf \nabla}\phi = 0$.
These lead to the boundary conditions $\dot c(0)=\dot c(t_f)=\dot
b_{\perp,\parallel}(0)=\dot b(t_f)_{\perp,\parallel}=0$ and $c(0)=1$,
$c(t_f)=\gamma_\perp^{14/5}\gamma_\parallel^{2/5}$,
$b_{\perp,\parallel}(0)=1$,
$b_\perp(t_f)=\gamma_\perp^{6/5}\gamma_\parallel^{-2/5}$ and
$b_\parallel(t_f)=\gamma_\perp^{-4/5}\gamma_\parallel^{8/5}$. These
latter imply that $\ddot b_{\perp,\parallel}(0)=\ddot
b_{\perp,\parallel}(t_f)=0$ must hold as well, giving sixteen
independent boundary conditions (BC).

Our procedure to engineer $\omega_{\perp,\parallel}(t)$ is to reduce
the dimensionality of the problem by only considering the trajectories that lead
to a constant axial size. This corresponds to keeping $b_\parallel (t)=b_\parallel(0)$ for any $t$, fixing a trap
decompression with $\gamma_\perp=\gamma_\parallel^2$.  In this case,
Eqs.~\eqref{eq:bperp}-\eqref{eq:a} reduce to
\begin{gather}
\label{eq:bperp1}
\ddot b_\perp(t)+ b_\perp(t)\omega_\perp^2(t) = 
\omega_{\perp}^2(0)/b_\perp^3(t) \\
\label{eq:bparallel1}
\omega_\parallel(t) = 
\omega_{\parallel}(0)/b_\perp(t)\\
\label{eq:a1}
b_\perp^4(t)\ddot c(t)+2b_\perp^3(t)
\dot b_\perp(t)\dot c(t)
+\omega_{\perp}^2(0) c(t)-\omega_{\perp}^2(0) b_\perp^3(t)=0.
\end{gather}
Equation~\eqref{eq:bperp1} is identical to Eq.~\eqref{eq:b} and
Eq.~\eqref{eq:a1} is nothing but Eq.~\eqref{eq:c(tau)} expressed 
with the real time (the rescaled time being given by Eq.~\eqref{eq:tau_bec} instead of Eq.~\eqref{eq:tau}).
Thus we can exploit for $b_\perp(t)$ and $c(t)$ the
solutions obtained for the non-interacting gas, provided that the
axial frequency is varied according to Eq.~\eqref{eq:bparallel1}.

\subsubsection{Example}
\label{sec:example_BEC}

As an example of the procedure described above, we determine the
trajectories used in Sec.~\ref{sec:experiment_bec} and in
Ref.~\cite{Schaff2010c}. The decompression parameters are given in
table~\ref{tab:param_bec}. The radial frequency is reduced by a factor
of 9, and the axial frequency by a factor of 3.

\begin{table}[ht]
\caption[Decompression parameters for the BEC.]{\label{tab:param_bec}Parameters of the 3D decompression of an interacting Bose-Einstein condensate.} 
\begin{indented}
\item[]\begin{tabular}{r|l}
Initial radial frequency $\omega_\perp(0)/2\pi$ & $235.8$~Hz \\
Final radial frequency $\omega_\perp(t_f)/2\pi$ & $26.2$~Hz \\
Initial axial frequency $\omega_\parallel(0)/2\pi$ & $22.2$~Hz \\
Final axial frequency $\omega_\parallel(t_f)/2\pi$ & $7.4$~Hz \\
Final rescaled time $\tau_f$       & $11.555$~ms \\
Corresponding duration $t_f$       & $30.0$~ms
\end{tabular}
\end{indented}
\end{table}

\begin{figure}[ht]
\includegraphics[width=\linewidth]{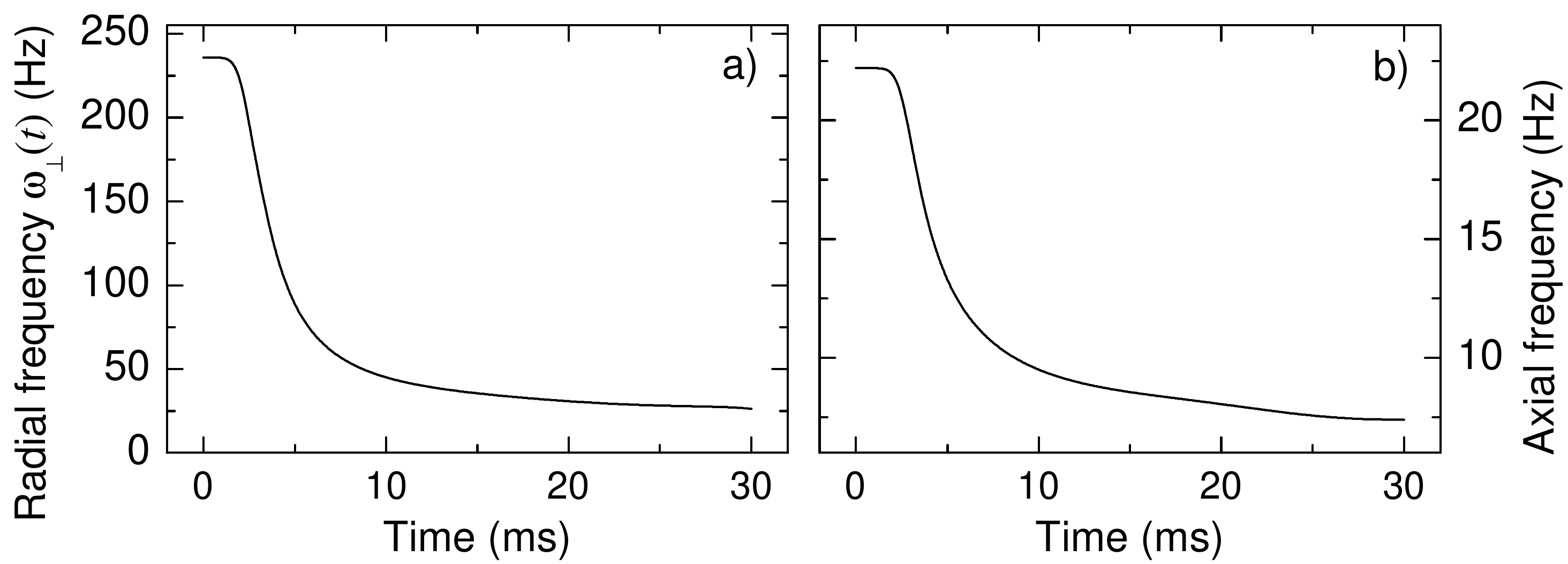}
\caption[Trap frequency trajectories for the shortcut decompression of
a Bose-Einstein condensate.]{(a) Radial and (b) axial trap frequencies
  for the shortcut decompression of a BEC in $t_f = 30$~ms.}
\label{fig:frequencies_bec}
\end{figure}

The obtained trajectories are represented in Fig.~\ref{fig:frequencies_bec}.

\subsubsection{Validity of the Thomas-Fermi approximation}

\begin{figure}[ht]
\includegraphics[width=\linewidth]{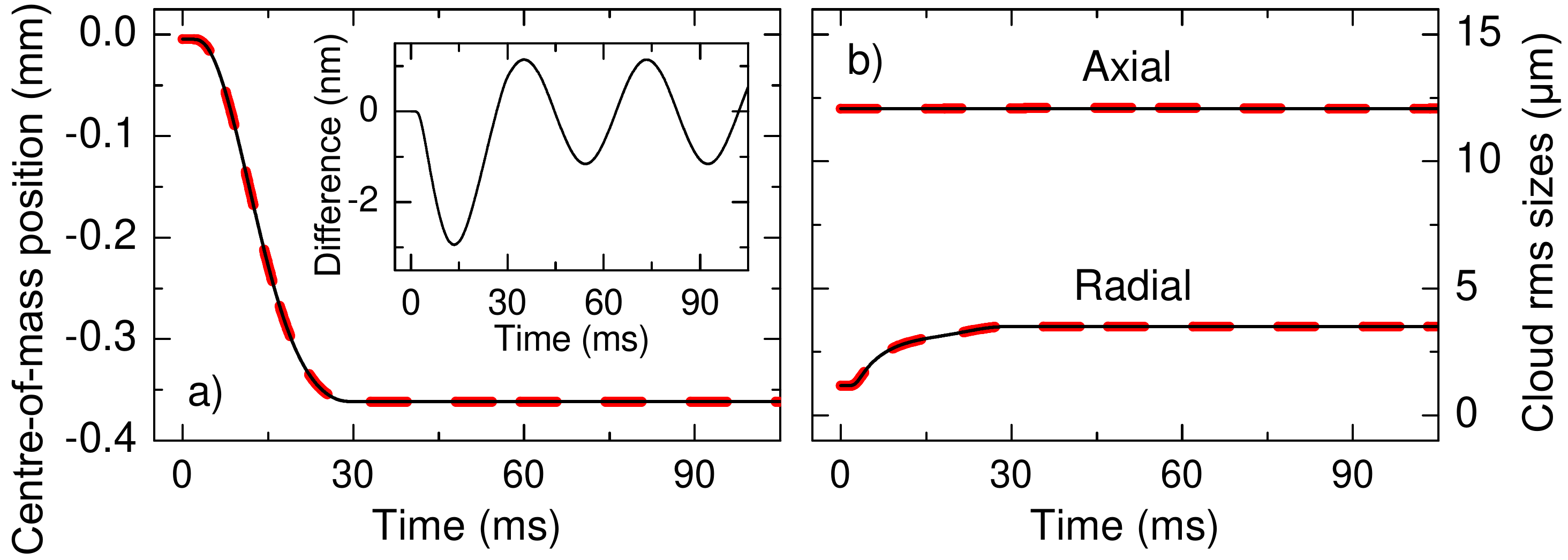}
\caption[Validity of the Thomas-Fermi approximation.]{Comparison
  between the GPE simulations (dashed red lines) and the scaling
  solutions given by the Thomas-Fermi approximation (solid black
  lines) showing its validity. (a) Centre-of-mass position; (b) axial
  and radial sizes. The peak relative difference between the scaling
  solution and the GPE simulations are respectively $0.3~\%$ and
  $0.2~\%$ for the axial and radial sizes. The decompression occurs during the first 30~ms, after which the cloud evolves in the static final trap.}
\label{fig:validity_TF}
\end{figure}

To check the validity of the Thomas-Fermi approximation that led to
the trajectories of Fig.~\ref{fig:frequencies_bec}, three-dimensional
Gross-Pitaevskii simulations have been performed and compared with the
analytical results of Sec.~\ref{sec:interacting_bec}. In the numerical
solution we use a split step operator in time combined with a fast
Fourier transformation in space. The results are presented in
Fig.~\ref{fig:validity_TF} and show that this approximation is well
justified for our experimental parameters (decompression of
Fig.~\ref{fig:frequencies_bec}, number of atoms $N \sim 10^5$,
scattering length of $^{87}$Rb of $a_s \sim 100 \, a_0$, $a_0$ being
the Bohr radius).

\section{Experimental realization of shortcuts to adiabaticity}
\label{sec:experiments}

The procedure described above was tested experimentally by quickly decompressing a trapped ultracold gas of $^{87}$Rb atoms. In this section, we describe the experimental steps involved in the preparation of the cold sample (cold thermal gas or BEC) and then explain how the decompression is controlled, monitored and compared to simpler (non-optimal) schemes.

\subsection{The apparatus}
\label{sec:setup}

\begin{figure}[ht]
\begin{center}
\includegraphics[width=0.7\linewidth]{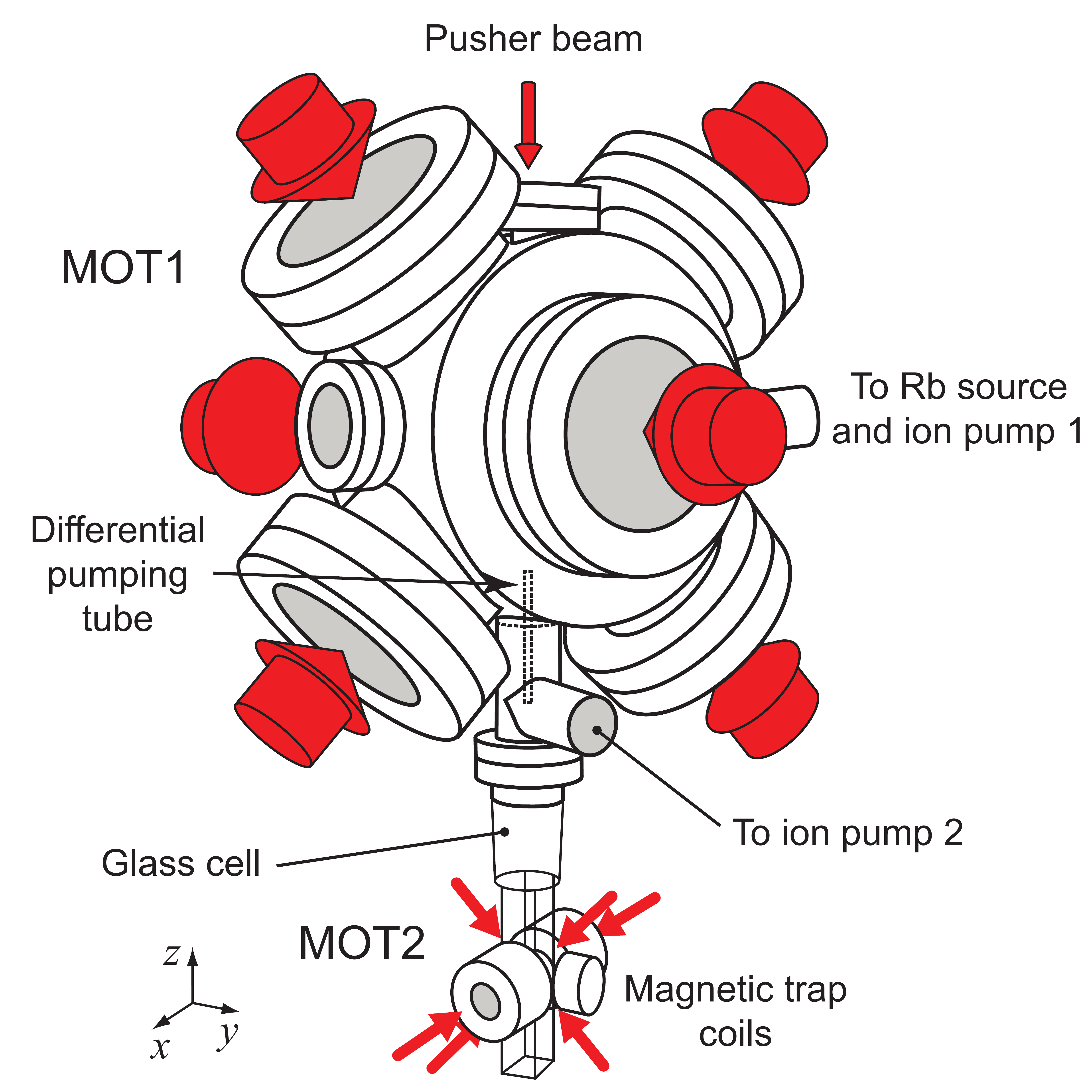}
\end{center}
  \caption[The apparatus.]{Schematic representation of the apparatus
    with the two magneto-optical trap chambers, the differential
    pumping tube, and the magnetic trap coils. The red arrows
    symbolize collimated laser beams.}
\label{fig:chambers}
\end{figure}

The Bose-Einstein condensation apparatus, sketched in
Fig.~\ref{fig:chambers}, is formed of two chambers connected by a
differential pumping tube. Each chamber is pumped by a separate ion
pump. A copper tube containing solid Rubidium provides gaseous
Rubidium to the upper chamber, resulting in a pressure of $\sim
10^{-9}$~mbar ($100$~nPa) which loads a large magneto-optical trap
(MOT1). The lower chamber is a glass cell, in which a second MOT can
be produced. The low-conductance tube connecting the chambers (length
10~cm, diameter 5~mm) results in a pressure on the order of
$10^{-11}$~mbar in the second chamber. This low pressure is essential
for the production of BECs because background gas collisions with the
magnetically trapped atoms is the key effect limiting the efficiency
of evaporative cooling.

\subsubsection{Production of ultracold clouds}

For the production of a BEC, the first step is the loading of MOT2
from MOT1. The light of both traps is red-detuned by $\delta = -
3.5~\Gamma$ from the $\ket{5^2 S_{1/2}, F=2} \rightarrow \ket{5^2
  P_{3/2}, F=3}$ cooling transition of $^{87}$Rb ($\Gamma/2\pi = 6.07$~MHz
is the natural line width of the $D_2$ transition of
$^{87}$Rb). The six beams of both traps also contain repumper light
tuned to the $\ket{5^2 S_{1/2}, F=1} \rightarrow \ket{5^2 P_{3/2},
  F=2}$ transition, which prevents atoms from accumulating in the
lowest energy state $\ket{5^2 S_{1/2}, F=1}$. The light is provided by
two DFB diode lasers, both injected in a single-pass tapered
amplifier. For both MOTs the light is delivered to the atoms by six
polarization-maintaining optical fibers, to ensure a good long-term
stability of the alignment. The total laser powers in MOT1 and 2 are
$360$~mW and $73$~mW respectively, with beam waists of $2.7$~cm and
$6.7$~mm. The magnetic field gradients of MOT1 and 2 are respectively
$B'_1 = 11$~G/cm and $B'_2 = 14.6$~G/cm (these values correspond to
the tighter axes).

While MOT2 is continuously on, MOT1 is operated in a pulsed
regime. First, the trapping light and magnetic field are on for 100~ms and
the MOT loads from the surrounding Rubidium vapour. Then, the light is
switched off and a blue-detuned \textit{pushing} laser beam, aligned
on the vertical axis linking the two MOTs and passing through the
differential pumping tube, is switched on for 6~ms. Because of the
radiation pressure force, this light pulse transfers the atoms
captured by MOT1 to MOT2 within $15~$ms. The force exerted by the
pushing beam is not strong enough to overcome the trapping force of
MOT2. After typically 5 to 10 seconds of such loading, MOT2 contains
$\sim 10^{10}$ atoms and MOT1 is then switched off (light and magnetic
field).  The cloud is then compressed by a temporal dark MOT
(compressed MOT): the cooling light detuning is changed from $\delta =
-3.5~\Gamma$ to $\delta = -8~\Gamma$ and the magnetic field gradient
is increased to $B'_2 = 65.5$~G/cm. This reduces the
multiple-scattering-induced repulsive interaction between atoms and
causes the cloud to shrink, thus increasing the density and collision
rate by a factor of $3$. The cloud is then further cooled to $80~\mu$K
by a 3-ms-long optical molasses phase (the field is switched off, and
the detuning changed to $\delta = - 10~\Gamma$).

For magnetic trapping, the atoms are then optically pumped to the
$\ket{5^2 S_{1/2}, F=2, m_F = 2}$ Zeeman substate by a beam detuned by
$\delta_\text{ZP} = + 3.2~\Gamma$ from the $\ket{5^2 S_{1/2}, F=2}
\rightarrow \ket{5^2 P_{3/2}, F=2}$ transition and a repumper beam,
detuned by $\delta_\text{ZP rep.} = - 3~\Gamma$ is applied to the
$\ket{5^2 S_{1/2}, F=1} \rightarrow \ket{5^2 P_{3/2}, F=2}$
transition. A homogeneous magnetic field of $\sim 0.5$~G is aligned
with the light wave vector to define the quantization axis. This
optical pumping stage lasts $300~\mu$s and then, all the light is
switched off and a quadrupole magnetic field ($54.1$~G/cm) is abruptly
turned on to trap the cloud.  This is followed by an adiabatic
compression of the cloud, performed by linearly increasing the
magnetic gradient to $278$~G/cm within $500$~ms. This compression
stage is primordial to increase the elastic collision rate to a high
enough value, an important requirement for evaporative cooling. At
this point, the number of atoms is $N \simeq 5 \times 10^9$, and the
temperature $T \simeq 400~\mu$K. In order to suppress Majorana losses,
the quadrupole magnetic trap is then converted into a Ioffe-Pritchard
trap by adiabatically ramping the current in a third coil
(quadrupole-Ioffe-configuration or QUIC trap~\cite{Esslinger1998}) within 500~ms. For cold
enough atoms, this anisotropic trap is harmonic with radial and axial
frequencies of $235.8$~Hz and $22.2$~Hz, respectively. Once the cloud
is in the Ioffe-Pritchard trap, radio-frequency (rf) evaporative
cooling is performed by ramping the rf frequency linearly from
$\nu_\text{start} = 20$~MHz to $\nu_\text{stop} = 1.3$~MHz within
$10$~s. We are able to produce almost pure BECs (no discernible
thermal fraction) containing up to $5\times10^5$ atoms. In order to
produce an ultracold thermal cloud, the loading time of MOT2 is
reduced to a few seconds. In this case, we are left, at the end of the
evaporation, with a dilute, thermal gas, with a low elastic collision
rate.

\subsubsection{Control of the trapping frequencies}

\begin{figure}[ht]
\begin{minipage}{0.5\linewidth}
\includegraphics[width=\linewidth]{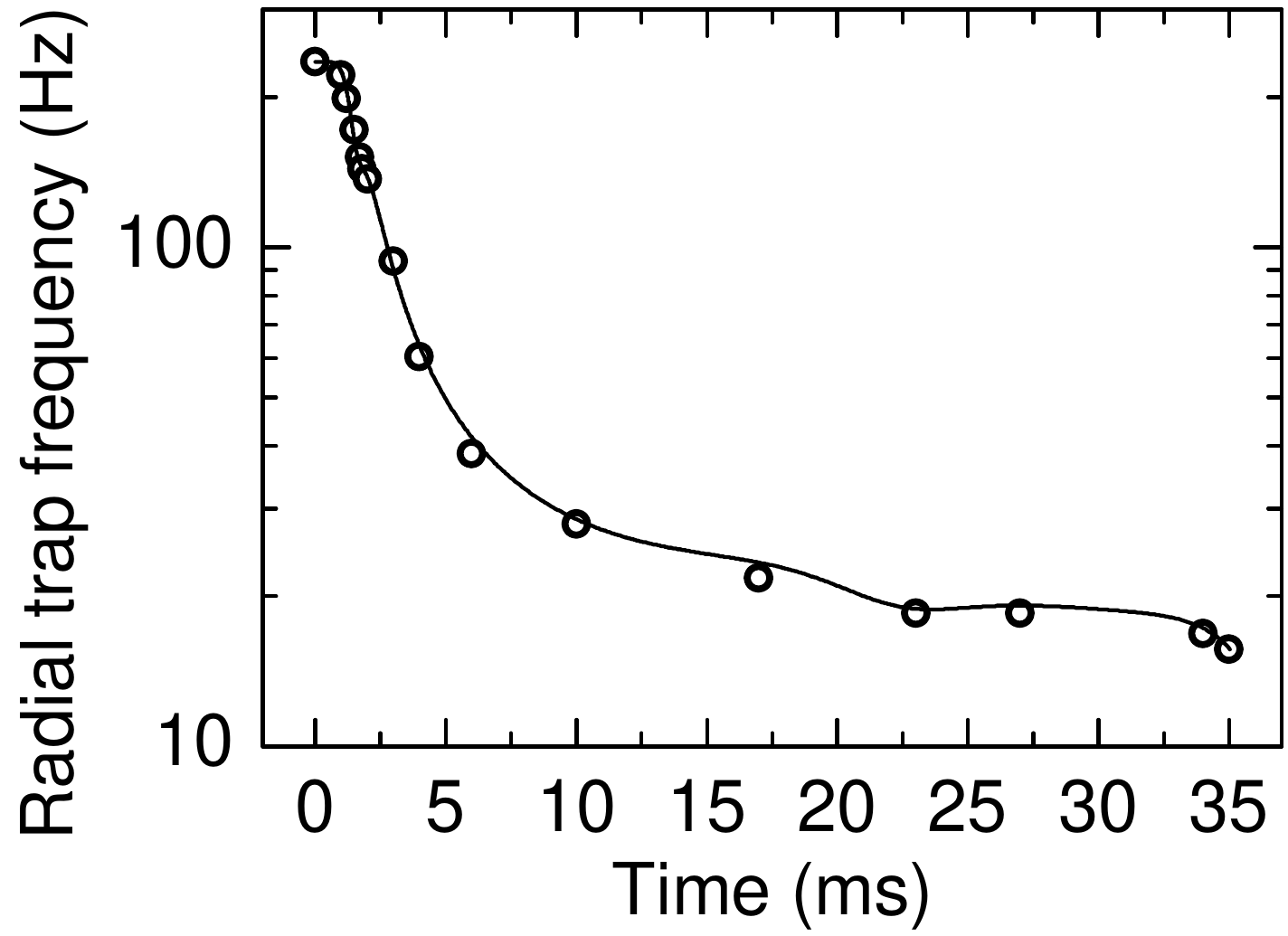}
\end{minipage} \hfill
\begin{minipage}{0.45\linewidth}
  \caption[Trap frequency calibration.]{Vertical trap frequency
    calibration. The solid line is the theoretical shortcut
    decompression trajectory, the circles are the measured
    frequencies. The parameters of the decompression are given in
    Tab.~\ref{tab:param_th}.}
\label{fig:trajectory_calibration}
\end{minipage}
\end{figure}

Implementing shortcuts to adiabaticity requires a precise control of
the trapping frequencies, in a dynamical fashion. In our QUIC magnetic
trap, this can be achieved by varying the current $i_Q$ running
through the 3 coils, and the current $i_{B_0}$ running through an
additional pair of Helmoltz coils disposed along the long (axial)
dimension of the trap (compensation coils). The resulting potential is
\begin{equation}
U(x,y,z) = \mu |\mathbf{B}| \simeq \mu \left[ B_0 + \frac{1}{2}\frac{B'^2}{B_0}\left(x^2 + z^2\right) + \frac{1}{2} B'' y^2 \right]
\label{eq:potential}
\end{equation}
where $\mu / h = 1.4$~MHz/G for atoms in $\ket{5^2 S_{1/2}, F=2, m_F =
  2}$. $B'$ is the radial magnetic field gradient while $B''$
corresponds to its curvature along $y$. The harmonic approximation of
Eq.~\eqref{eq:potential} describes accurately the potential seen by cold
enough atoms i.e. $k_B T \ll \mu B_0$~\cite{Foot2002}. Then, the
radial and axial angular frequencies are
\begin{equation}
\omega_\perp \defeq \omega_z \simeq \omega_x  \simeq \sqrt{\frac{\mu}{m}} \frac{B'(i_Q)}{\sqrt{B_0(i_Q, i_{B_0})}} , \quad
\omega_\parallel \defeq \omega_y = \sqrt{\frac{\mu}{m}}\sqrt{B''(i_Q)} .
\label{eq:frequencies}
\end{equation}
These expressions show that the radial and axial frequencies can be
controlled independently to some extent. The dynamical control of the
currents is achieved using homemade, computer-controlled electronic
circuits. The experimental realization of shortcut trajectories
requires a careful preliminary calibration of the frequencies versus
currents, which is achieved by monitoring the cloud's centre-of-mass
oscillations after a small excitation. Due to the finite time response
of the controlling circuit, it is also necessary to check the behavior
of the frequency during an actual trajectory. This is illustrated in
Fig.~\ref{fig:trajectory_calibration}, where we compare the
theoretical decompression trajectory of Fig.~\ref{fig:c_b_t_nu} (line)
and measured experimental values (circles). In this example, the
deviation is below 5$\%$.

\subsection{Shortcut to adiabaticity for a non-interacting gas}
\label{sec:experiment_th}

For this first experiment, we use the procedure described in
Sec.~\ref{sec:setup} to produce a thermal gas ($N \simeq 10^5 $, $T_0 = 1.6~\mu$K) in
the compressed trap of frequencies $\omega_x(0)/2\pi = 228.1$~Hz,
$\omega_y(0)/2\pi = 22.2$~Hz and $\omega_z(0)/2\pi = 235.8$~Hz. The initial cloud-averaged collision rate per particle is $\gamma_\text{el} \simeq 8$~Hz, which corresponds to a collision time of $\sim 125$~ms. This is 30 times the oscillation period, and more than 3 times the decompression time, which justifies the non-interacting approximation.

We use here the decompression trajectory discussed in
Sec.~\ref{sec:example_thermal}, adapted to the vertical axis $(Oz)$,
with the parameters of Tab.~\ref{tab:param_th}. To maximize the
decompression factor $\gamma^2 = \omega_z(0)/\omega_z(t_f)$, the
compensation coils current $i_{B_0}$ is increased from $i_{B_0}(t=0)
\simeq 0$~A to $i_{B_0}(t_f) = 3.0$~A, while the QUIC current is
decreased from $i_{Q}(t=0) = 26.7$~A to $i_{Q}(t_f) = 3.6$~A (see the
resulting trajectory in Fig.~\ref{fig:trajectory_calibration}). The
decompression duration is $t_f = 35$~ms.

In theory, starting from a gas at equilibrium and temperature $T_0$ in
the compressed trap, a shortcut to adiabaticity should lead to an
equilibrium state in the final trap, with a temperature $T_f = T_0 \,
\omega(t_f)/\omega(0)$. This corresponds to a situation where entropy
has not increased. On the contrary, for a non-optimal decompression,
one expects to observe oscillations of the cloud's size and centre of
mass in the decompressed trap, once the decompression is completed. To
evaluate the efficiency of our shortcut, we thus perform the fast
decompression, and hold the cloud in the decompressed trap for a
variable amount of time. The trap is then abruptly switched off, and
an absorption image is taken after a constant time of free expansion
($6$~ms). The amplitude of the dipole (oscillation of the centre of
mass) and breathing modes (oscillation of the size) give access to the
excess energy provided to the cloud, as compared to an adiabatic
modification of the potential. If the cloud is reasonably at
equilibrium after decompression, one can also directly measure the
final temperature by measuring the evolution of the size during a free
expansion.

In the following, we compare four decompression trajectories:
\begin{enumerate}
\item the shortcut, given on Figs.~\ref{fig:c_b_t_nu}d and \ref{fig:trajectory_calibration},
\item a linear decompression of the same duration (35~ms),
\item an abrupt decompression, which, somehow, corresponds to a worst case scenario (in practice, the decompression time is 0.1~ms and $\omega(t)$ is not controlled, it is imposed by the response of the magnetic trap control electronics),
\item a 6-s-long linear decompression, which can be considered nearly adiabatic.
\end{enumerate}
What is referred to as `linear decompression' corresponds to both
control currents being varied linearly with time. The corresponding
frequency trajectory \emph{is not linear}.

\begin{figure}[ht]
\begin{center}
\includegraphics[width=\linewidth]{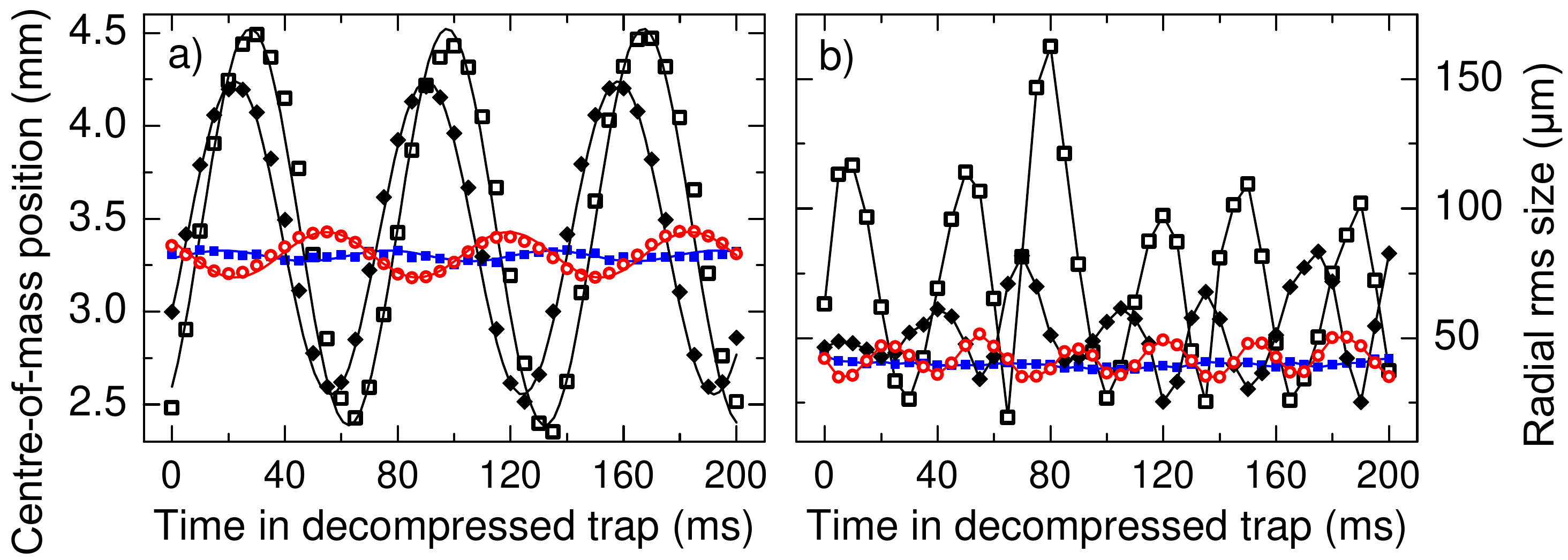}
\caption[Measures of the cloud size and centre-of-mass position for
different decompression schemes.]{Comparison between different trap
  decompression schemes (along the vertical axis). Open red circles:
  shortcut decompression in 35~ms; black diamonds: linear
  decompression in 35~ms; solid blue squares: linear decompression in
  6~s; open black squares: abrupt decompression. The decompression is
  performed, and then, the cloud is held in the decompressed trap for
  a variable time. We monitor (a) the vertical centre-of-mass position
  (dipole mode) and (b) the cloud size (breathing mode), after 6~ms
  time of flight. On (a), the solid lines are sine fits, on (b) they
  just connect the points to guide the eye.}
\label{fig:xp_position_size}
\end{center}
\end{figure}

The experimental results are summarized on
Fig.~\ref{fig:xp_position_size}. In the case of the 6-s-long linear
ramp (filled squares), very little residual excitation is observed
(although the residual dipole mode is still measurable). In the
shortcut case (open circles), clear oscillations of the cloud width
and centre-of-mass position are seen, but they are much reduced
compared to the fast linear ramp (diamonds) and abrupt decompression
(open squares).

Compared to the linear decompression in 35~ms, the shortcut reduces
the amplitude of the dipole mode by a factor of $7.2$ (obtained from
the sine fits) and the amplitude of the breathing mode by a factor of
$3$ (comparison of the standard deviations of the two sets of
data). The excess energy, which is dominated by the centre-of-mass
energy, is thus reduced by a factor of $\sim 52$. In the case of the
6-s-long ramp, we measured a final temperature of the cloud of
$130$~nK, a factor 12.5 below the initial one. This is consistent with
the expected value of 15. The small difference may arise from a small
heating rate due to the fluctuations of the magnetic trap.

The fact that the shortcut decompression still produces sizeable
excitations is due to experimental imperfections. Several possible
causes can be invoked. Firstly, as seen on
Fig.~\ref{fig:trajectory_calibration}, there are still small
deviations from the ideal trajectory. These may have an impact,
especially in the last phase of the trajectory where the cloud is
subject to a large acceleration (see
Fig.~\ref{fig:th_position_size}). Second, as can again be seen in
Fig.~\ref{fig:th_position_size}, during the trajectory the cloud
wanders quite far (several hundred $\mu$m) from the trap centre and
feels the non-harmonic part of the potential. This effect is difficult
to quantify since our knowledge of the potential shape is not
sufficiently accurate (however, the anharmonicity could be inferred
from variations of the oscillation frequency with amplitude). In
principle, it could be avoided by designing other shortcut
trajectories keeping the cloud closer to the trap centre at all times.

\begin{figure}[ht]
\begin{center}
\begin{minipage}{0.51\linewidth}
\includegraphics[width=\linewidth]{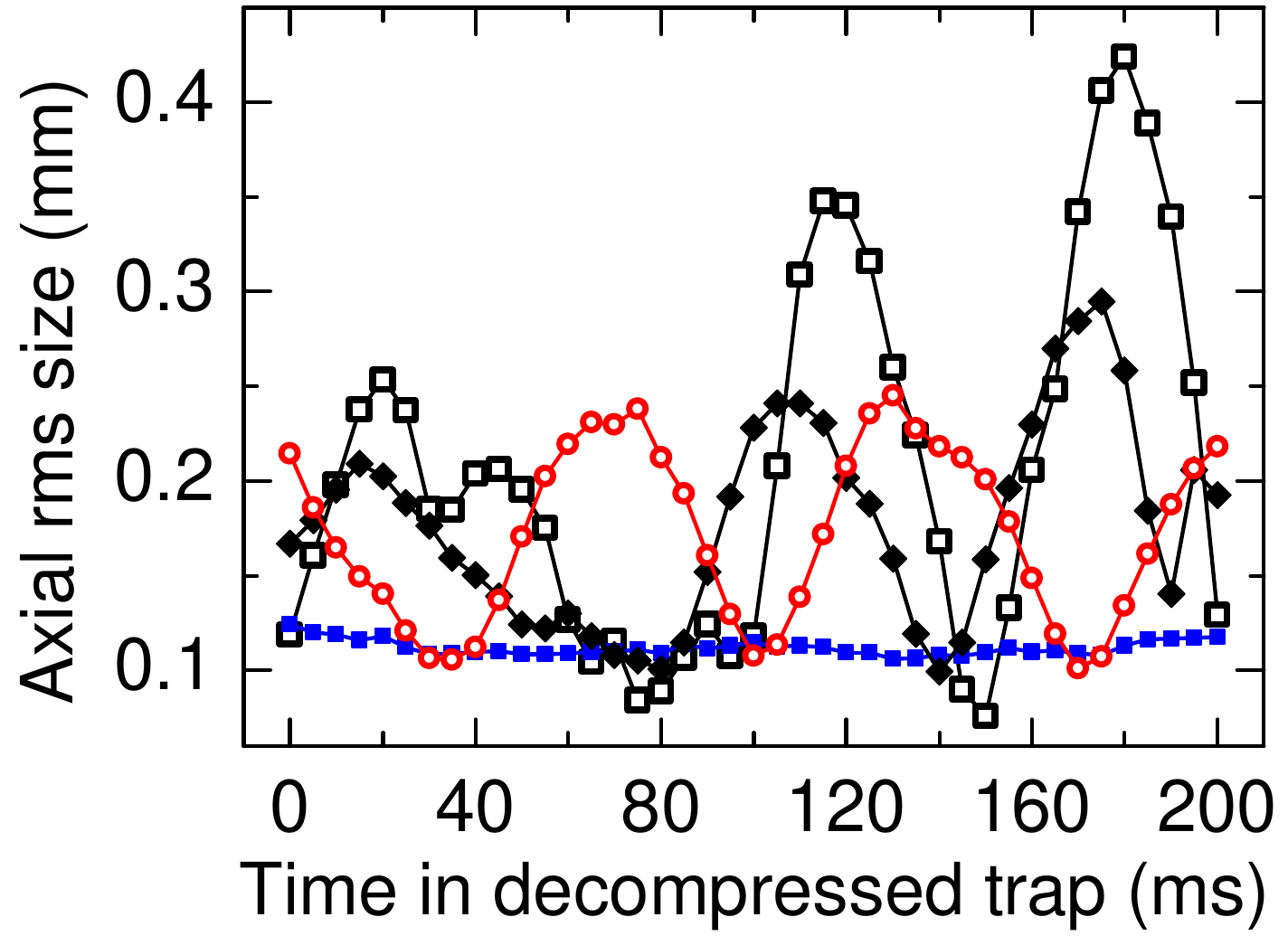}
\end{minipage} \hfill
\begin{minipage}{0.45\linewidth}
  \caption[Measures of the axial cloud size.]{Impact of the vertical
    decompression schemes on the axial size ($y$ direction). Same
    colors and symbols as in Fig.~\ref{fig:xp_position_size}. The
    amplitude of the axial breathing mode is not affected by the use
    of a shortcut trajectory adapted to the radial dimensions.}
\label{fig:xp_sy}
\end{minipage}
\end{center}
\end{figure}

Fig.~\ref{fig:xp_sy} shows the behavior of the axial size of the cloud
in the conditions of Fig.~\ref{fig:xp_position_size}b. Since the
shortcut trajectory was designed only for the radial dimensions, the
resulting axial breathing mode is of the same magnitude as for the
linear decompression.

\begin{figure}[ht]
\begin{center}
\includegraphics[width=0.5\linewidth]{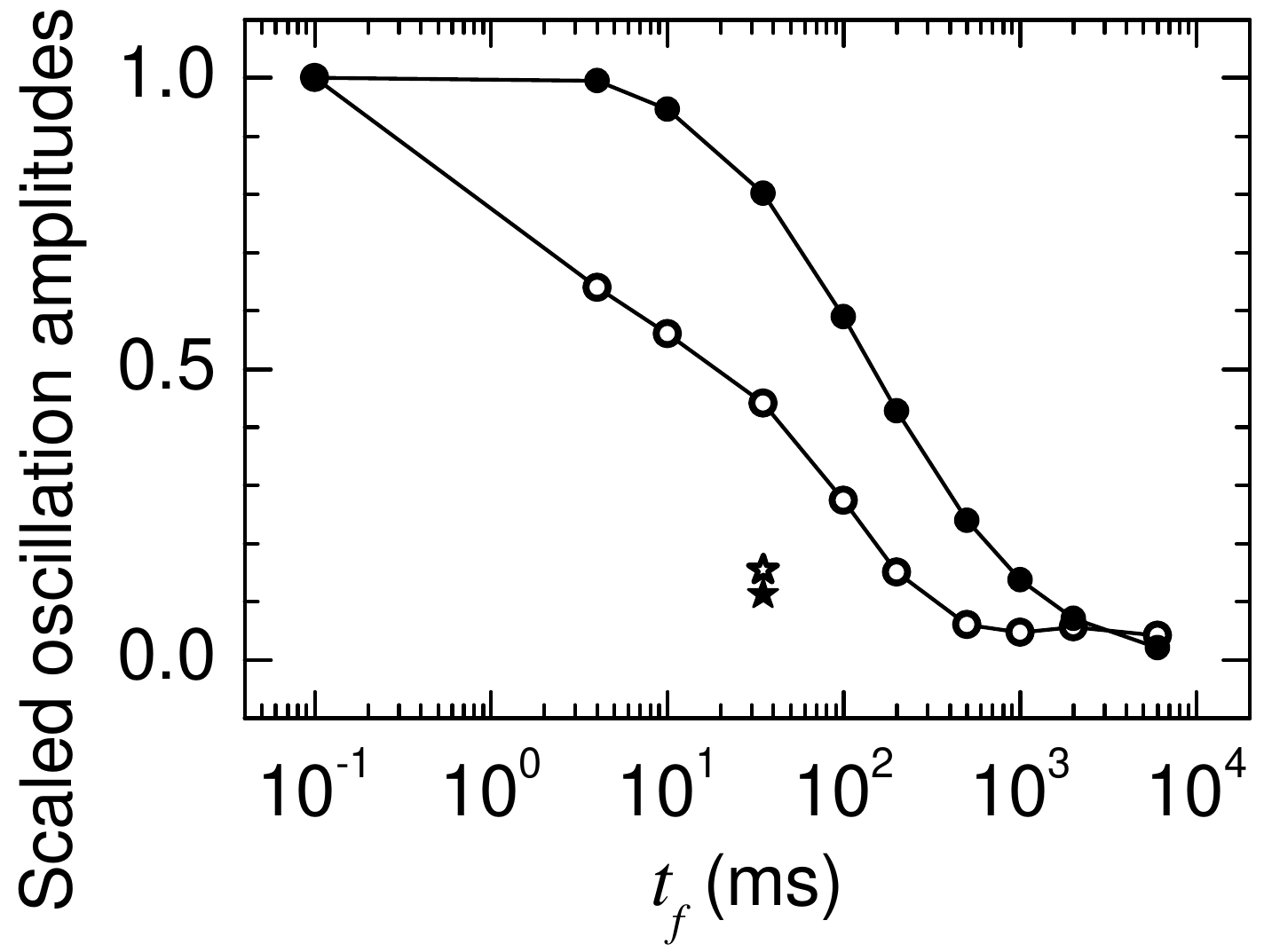}
\caption[Adiabaticity.]{Comparison between linear and shortcut
  decompression schemes. We plot the scaled oscillation amplitudes of
  the breathing (cloud size, open symbols) and dipole (centre-of-mass
  position) modes versus the decompression time $t_f$. The circles and
  stars correspond to linear and shortcut decompressions,
  respectively.}
\label{fig:adiabaticity}
\end{center}
\end{figure}

We compare on Fig.~\ref{fig:adiabaticity} the results of the shortcut
decompression to linear ramps of various durations. The vertical axis
in this figure represents amplitudes of oscillations after trap
decompression, either of the centre-of-mass position (filled symbols)
or of the cloud radius (open symbols), scaled by their values for an
abrupt decompression ($t_f \sim 0.1$~ms). The horizontal axis is the
duration of the decompression $t_f$ (notice the logarithmic scale). The
circles correspond to linear decompressions while the stars are the
shortcut results. As can be seen, fulfilling the adiabaticity
criterion is easier for the breathing mode (size oscillation) than for
the dipole mode (centre-of-mass oscillation): the oscillation
amplitude is reduced by a factor of $2$ for $t_f = 20$~ms for the
earlier, and for $t_f \simeq 150$~ms for the latter. Using the
amplitude of the dipole mode as a criterion to compare the linear and
shortcut schemes, one sees that the decompression time is reduced by a
factor of 37.

\subsection{Shortcut to adiabaticity for an interacting condensate}
\label{sec:experiment_bec}

As opposed to the previous case of non-interacting atoms, the
decompression of a BEC is an intrinsically 3D problem because of the
interactions. As a result, both the radial and axial frequencies have
to be varied following Eqs.~\eqref{eq:bperp1} and
\eqref{eq:bparallel1} in order to realize a shortcut to
adiabaticity. In the present section, we describe a decompression
experiment based on the trajectories discussed in
Sec.~\ref{sec:example_BEC} and represented in
Fig.~\ref{fig:frequencies_bec}. In this scheme, the radial frequency
is decreased by a factor of 9, while the axial frequency is adjusted to
maintain the \emph{axial size} of the BEC \emph{fixed} during the
whole trajectory. Accordingly, the axial frequency is decreased by a
factor of 3.

We start from an initial BEC containing $1.3 \times 10^5$ atoms in the
condensed fraction, and $7 \times 10^4$ non-condensed atoms at a
temperature of 130~nK. The experimental scheme is similar to that
employed for the thermal cloud. Here, we use a longer time of flight
of 28~ms to characterize the various excitations generated by rapid
decompressions. Three decompression schemes are compared:
\begin{enumerate}
\item the shortcut to adiabaticity in 30~ms,
\item the linear decompression in 30~ms,
\item an abrupt decompression.
\end{enumerate}
Contrary to the previous case of a thermal cloud, the BEC cannot be
held for more than 150~ms in the compressed magnetic trap because of
a relatively high heating rate. Thus, here we cannot compare our
scheme to the adiabatic limit corresponding to a slow linear
decompression.

\begin{figure}[ht]
\begin{center}
\includegraphics[width=0.8\linewidth]{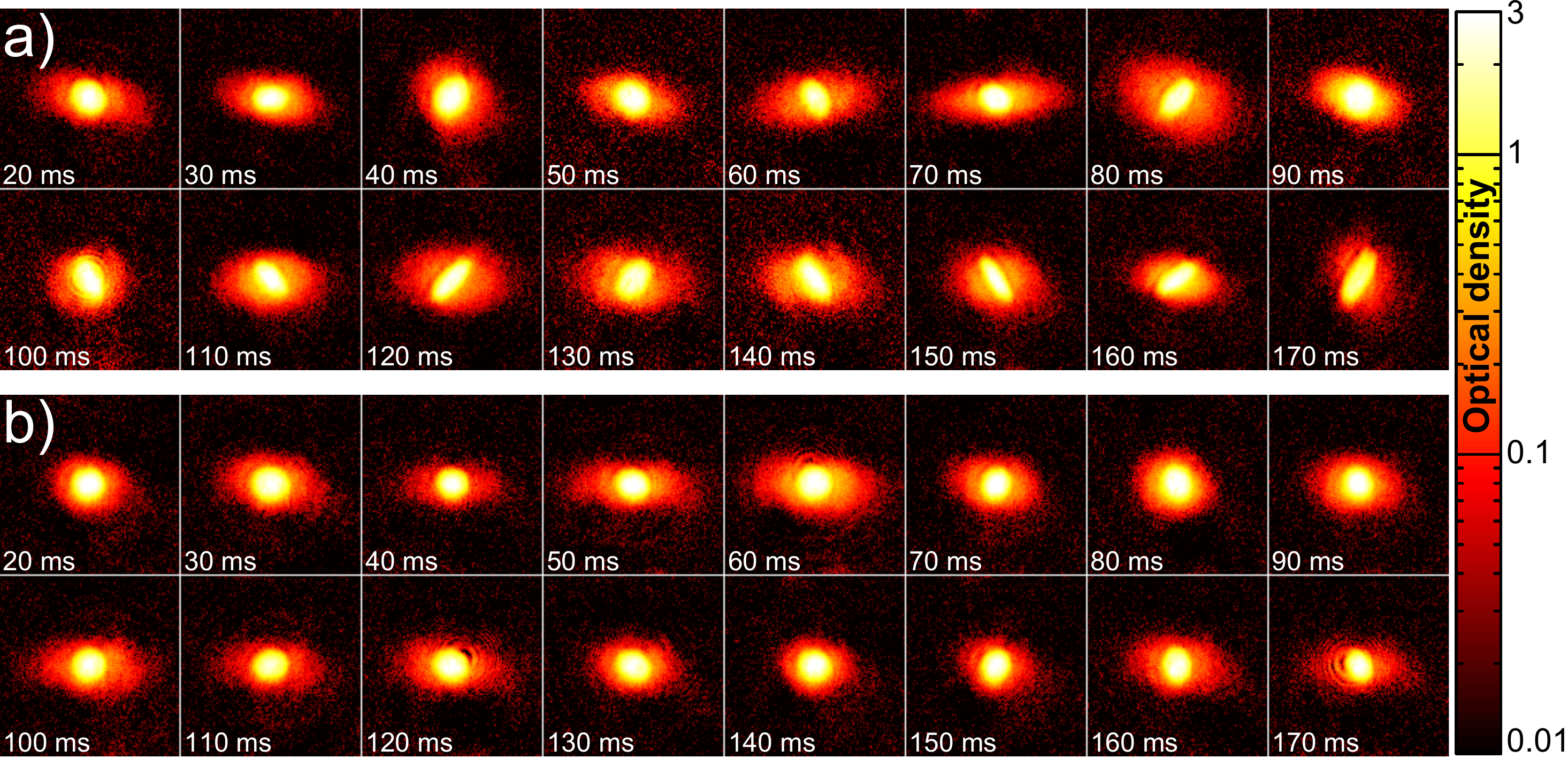}
\end{center}
\caption[Fast decompression of a Bose-Einstein condensate. Comparison of optimal and non-optimal schemes.]{Comparison of linear and shortcut BEC decompressions. We compare the time evolution of the BEC after two different decompression schemes: (a) a 30-ms-long linear ramp and (b) the shortcut trajectory (see text). The centre-of-mass motion has been subtracted from these time-of-flight images for clarity. On each image, the region where the optical density is highest (yellow and white) correspond to the condensate, while the red halo is the thermal component.}
\label{fig:xp_bec_images}
\end{figure}

Figure~\ref{fig:xp_bec_images} shows the temporal behaviour of the
cloud following the linear and shortcut decompressions. These
absorption images are taken in the ($y, z$) plane, after a certain
holding time in the decompressed trap (indicated in the figure) plus a
28-ms-long time of flight. The field of view is $545\,\mu\mathrm{m} \times
545\,\mu$m. The centre-of-mass motion has been subtracted from these
data for better clarity. In the linear case the BEC (yellow central part) experiences large
deformations and oscillations of its aspect ratio, whereas in the
shortcut case it remains nearly perfectly stationary. Surprisingly, in
the case of the linear decompression the BEC also oscillates
\emph{angularly}. We attribute this to an uncontrolled tilt of the
trap axes during the decompression. This will be discussed in more
details later. The nearly isotropic aspect of the BEC after the
shortcut decompression is due to the value of the time of flight,
which is close to the critical time of inversion of the aspect ratio.
The thermal component surrounding the BEC (red halo) is also visible. 
It's temporal evolution is discussed at the end of this section.

To provide a more quantitative analysis, the column densities obtained from the absorption images were 
fitted with a 2D bimodal distribution consisting of a Gaussian component, accounting for the thermal fraction, plus a 3D
inverted parabola integrated along one dimension, accounting for the condensed atoms. The fitting parameters were the cloud centre, two angles, one for each couple of eigenaxes of each components, and the two widths of each components.

\begin{figure}[ht]
\includegraphics[width=\linewidth]{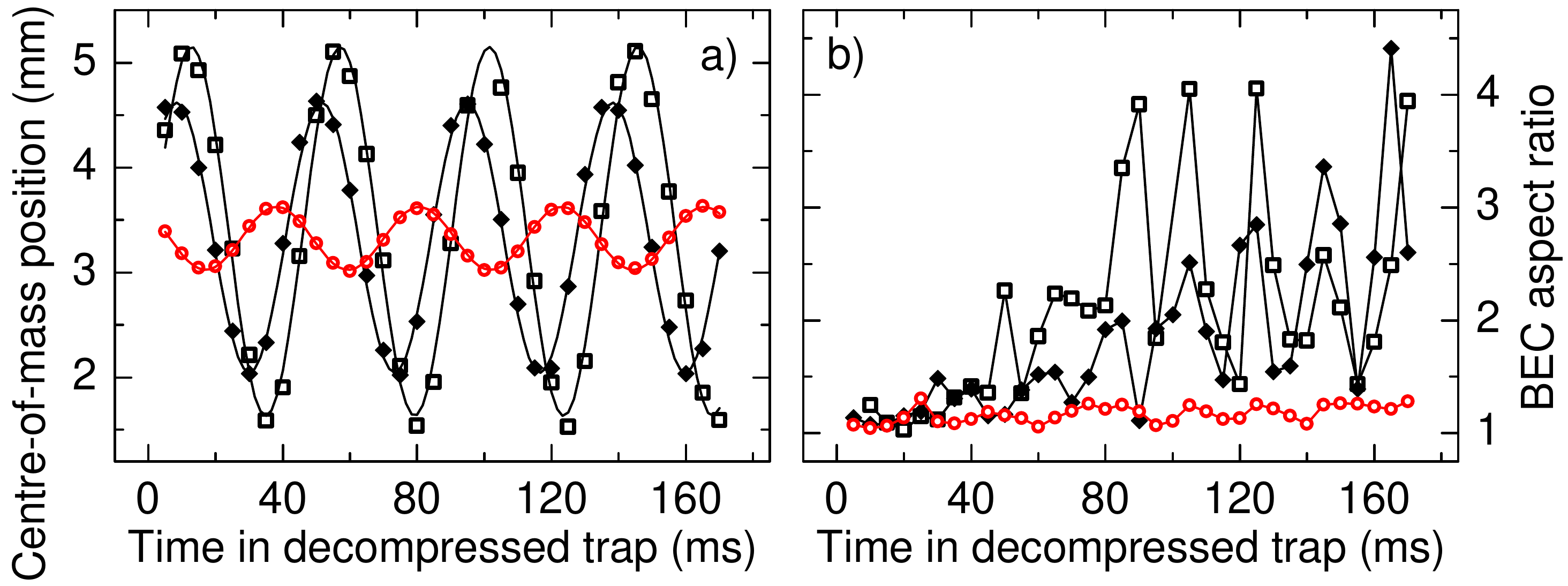}
\caption[Excitations induced by the fast decompression of a
Bose-Einstein condensate.]{Decompression-induced excitations of the BEC. We
  report the temporal evolution of (a) the centre-of-mass position and
  (b) the aspect ratio of the BEC after three different decompression
  schemes: an abrupt decompression (black squares); a 30 ms linear
  ramp (black diamonds); the 30 ms shortcut trajectory (red
  circles). All measurement are performed after 28~ms of time of flight.}
\label{fig:xp_position_size_bec}
\end{figure}

In Fig.~\ref{fig:xp_position_size_bec}a) is reported the
centre-of-mass oscillations (dipole mode) for the abrupt (squares),
linear (diamonds) and shortcut
(circles). Figure~\ref{fig:xp_position_size_bec}b) shows the
oscillations of the BEC's aspect ratio (breathing mode). All
measurements are performed after a 28~ms time of flight. As in the
case of the non-interacting cloud, the shortcut scheme reduces the amplitude
of the dipole mode compared to a standard linear decompression, here
by a factor of 4.3. For our relatively long time of flight, the measured
positions reflect the atomic velocities. Thus, using the shortcut
scheme reduces the kinetic energy associated with the dipole mode by a
factor of $18.5$ compared to the linear one (and 36 compared to the
abrupt). The residual energy after the shortcut decompression is
580~nK. As can be seen in Fig.~\ref{fig:xp_position_size_bec}b), both
non-optimal schemes induce very large oscillations of the BEC's aspect
ratio, with a rather complicated dynamics. A Fourier analysis reveals
a main oscillation frequency of 47 Hz, consistent with a radial
breathing mode at $2 \omega_{\perp}$~\cite{Mewes1996, Chevy2002,
  Dalfovo1999}. A smaller contribution at 12.5 Hz corresponds to the
axial breathing mode at $\sqrt{5/2}
\omega_{\parallel}$~\cite{Dalfovo1999}. The shortcut scheme suppresses
strikingly these breathing oscillations, yielding a BEC close to the
targeted equilibrium state.

\begin{figure}[ht]
\includegraphics[width=\linewidth]{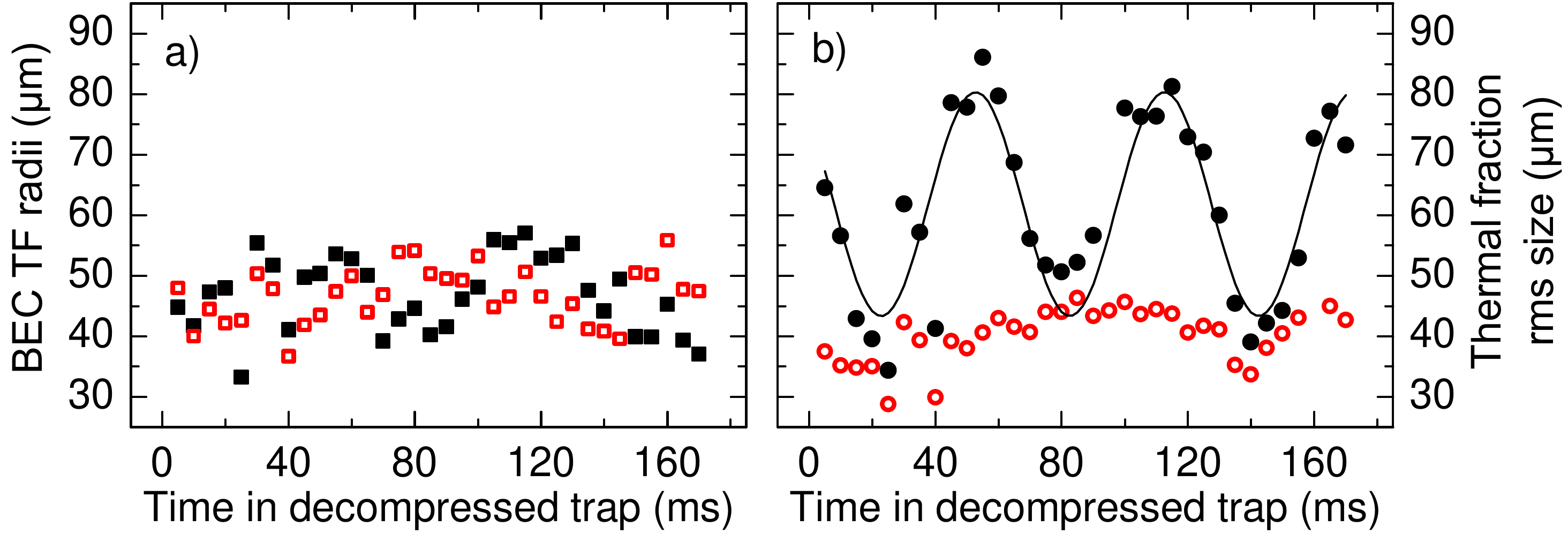}
\caption[Comparison of the decompression-induced excitations of the
condensed and thermal fractions.]{BEC versus thermal cloud
  decompression. We plot (a) the sizes of the BEC and (b) thermal
  component versus the time spend in the decompressed trap for the
  shortcut trajectory. The filled and empty symbols correspond to the
  axial and radial (vertical) directions respectively. The line is a
  sine fit to the thermal fraction axial size.}
\label{fig:xp_size_thermal_bec}
\end{figure}

As emphasized in section~\ref{sec:interacting_bec}, the shortcut
trajectory employed in this experiment is also valid for the thermal
fraction, in the radial dimensions only. This is demonstrated in
Fig.~\ref{fig:xp_size_thermal_bec}, where we compare the oscillations
of the radial (open symbols) and axial (filled symbols) sizes of a)
the BEC, and b) the thermal fraction, after the shortcut
decompression. The BEC's TF radius is stationary with an average value
of $46.8~\mu$m close to the theoretical value ($43 \mu$m). As can be
observed in Fig.~\ref{fig:xp_size_thermal_bec}b), the radial size of
the thermal fraction is also quite stationary as expected from a
shortcut trajectory. Thus, this experiment demonstrates that both a
non-interacting thermal gas and an interacting BEC can be decompressed
simultaneously using an appropriate shortcut trajectory. The observed
behavior is also qualitatively consistent with our initial assumption
that the BEC and thermal fraction
are independent. However, we expect that ultimately the validity of
this approach will be limited by the interaction between the condensed
and non-condensed fractions. The temperature inferred from the radial
size of the thermal component is 22~nK, a factor of 6 below the
initial one. This factor is smaller than the expected one
($\omega_\perp(0)/\omega_\perp(t_f)= 9$), and even if we improve the
experimental set-up to realize the ideal frequency trajectory we would
probably be limited by the transfer of energy from the axial breathing
mode {\it via} the interaction with the condensate.  Indeed, the axial
size of the thermal fraction presents clear breathing oscillations,
reflecting the fact that the shortcut trajectory
$\omega_{\parallel}(t)$ is not valid in this case, as expected.

\begin{figure}[ht]
\begin{center}
\includegraphics[width=0.51\linewidth]{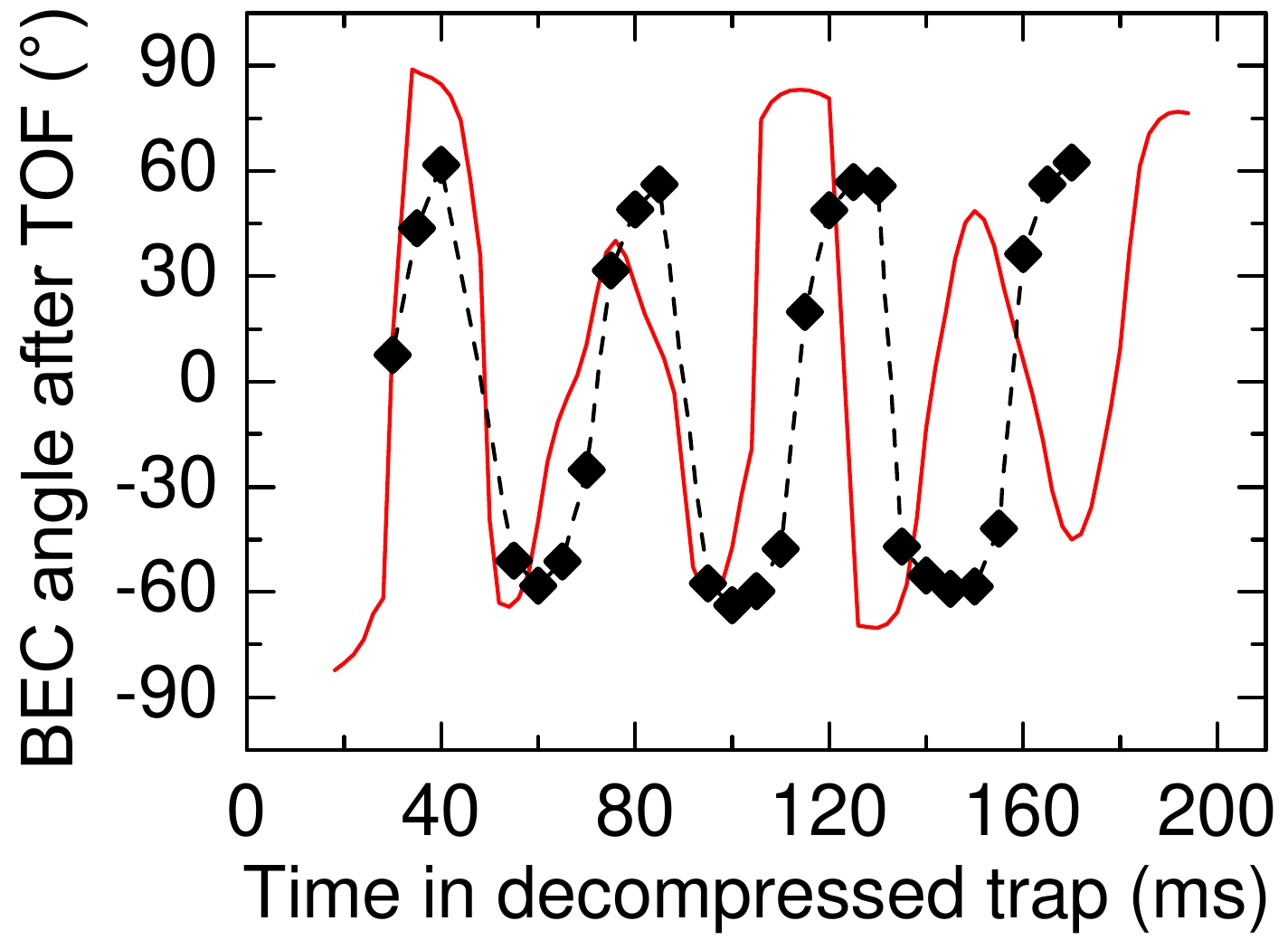}
\end{center}
\caption[Scissors mode excitation.]{Experimental observation of a
  scissors mode excitation following the linear decompression
  (diamonds). The red line is a GPE simulation. The oscillation is not
  quantitatively reproduced because it depends on the precise way the
  trap is rotated during decompression, which is not known for the
  whole trajectory. Only the final tilt of $3^\circ$ was measured. For
  the GPE simulation, the trap angle was assumed to be proportional to
  the trap bottom displacement from its original position.}
\label{fig:scissors}
\end{figure}

A striking feature in Fig.~\ref{fig:xp_bec_images}a was the large
angular oscillation of the BEC after the linear decompression. This
unexpected effect is due to a slight tilt of the QUIC trap
eigenaxes ($3^\circ$) in the $(y, z)$ plane as the trap centre moves downwards due
to gravity. Because of this, an angular momentum is imparted to the
atoms during the decompression, exciting a `scissors
mode'~\cite{Stringari1999,Marago2000}. Our nearly critical time of flight 
then results in a magnification and a deformation of the
scissors oscillations~\cite{Edwards2002,
  Modugno2003}. Fig.~\ref{fig:scissors} shows an example of these
oscillations, together with a GPE simulation (red line).

\section{Other possible applications}

In this section, we attempt to generalize the shortcut decompression of Bose-Einstein condensates to other situations which may find applications in experiments where a fast and large modification of the width of the velocity distribution or of the chemical potential is required.

\subsection{Arbitrary variation of a harmonic potential}
Let us consider the time evolution of a condensate in the 
time-dependent harmonic potential of the form
\begin{equation}
U(\mathbf{r},t) = \dfrac{1}{2}m \, \mathbf{r}^t W(t) \mathbf{r} + \mathbf{r}^t \boldsymbol{u}(t)
\label{generalpot}
\end{equation}
where the symmetric matrix $W(t)=R^{-1}(t) \tilde W(t) R(t)$
represents the harmonic potential of stiffness
\begin{equation}
\tilde W=\left(\begin{tabular}{ccc}
$\omega_x^2(t)$ &0&0\\
0&$\omega_y^2(t)$&0\\
0&0&$\omega_z^2(t)$\\
\end{tabular}
\right),
\end{equation}
rotated by a rotation matrix $R(t)$. The column vectors
$\mathbf{r}$ and $\boldsymbol{u}$ respectively represent the
position and a spatially homogeneous force which may depend on
time. The superscript $^t$ indicates the transpose of vectors or
matrices.

To solve Eq.~\eqref{eq:gpe} we look for a linear change of variables
$\boldsymbol{\rho}(\mathbf{r}, \{b_{ij}(t)\},
\{r_i^{\text{cm}}(t)\})$ where the $b_{ij}$'s are scaling and rotation
functions for the $r_i$'s. Let $B$ be a 3$\times$3 matrix which
elements are the functions $b_{ij}$. The transformation is
\begin{equation}
\boldsymbol{\rho}=B^{-1}(t) \left(\mathbf{r}-\mathbf{r}^{\text{cm}}(t)\right) = B^{-1}(t)\mathbf{r}+\boldsymbol{a}(t) .
\end{equation}
In the TF limit, and if the matrix $\dot{B}B^{-1}$ is symmetric, Eq.~\eqref{eq:gpe} is
invariant under this transformation. The full derivation is given in \ref{sec:appA}, but we give here the key elements.

The TF approximation
consists in neglecting the kinetic-energy-like term 
\begin{equation}  
\sum_{i,j,k}[B^{-1}]_{ij}[B^{-1}]_{kj}
\dfrac{\partial^2\chi}{\partial\rho_i\partial\rho_k},
\label{brutto}
\end{equation}
$\chi(\boldsymbol{\rho},\tau)$  being defined as in Eq. \eqref{defchi}.
In this regime, the condensate wavefunction $\chi(\boldsymbol{\rho},\tau)$ 
verifies the equation of motion Eq.~\eqref{eq:gpe2}, under the action of the 
time-independent potential
\begin{equation}
U(\boldsymbol{\rho}, 0)=\dfrac{1}{2}m\boldsymbol{\rho}^tW(0)\boldsymbol{\rho}
\label{potsemplice}
\end{equation}
if the generic scaling functions satisfy
\begin{equation}
\ddot{B}^tB+B^tWB=\dfrac{W(0)}{{\rm det}B},
\label{scaling_full}
\end{equation}
\begin{equation}
\ddot{\mathbf{r}}^{\text{cm}} + 
W(t)\boldsymbol{r^{\text{cm}}} -\dfrac{1}{m}\boldsymbol{u} = \boldsymbol{0}.
\label{shift_full}
\end{equation}
It is worthwhile recalling that, as shown by the above equations, the
evolution of $B$ is decoupled from the centre-of-mass motion which
evolves with the net external force.  The phase of the wavefunction is
chosen as
\begin{equation}
\phi(\mathbf{r},t)={\frac{m}{\hbar}}\left\{\dfrac{1}{2}\mathbf{r}^t
\dot{B}B^{-1}\mathbf{r}-\mathbf{r}^tB\dot{\boldsymbol{a}}\right\}+
\phi_0(t),
\label{phase_gen}
\end{equation}
with
\begin{equation}
\phi_0=-\frac{m}{2\hbar}\int_0^t {\rm d}t'\left(\dot{\boldsymbol{a}}^tB^tB
\dot{\boldsymbol{a}}-\boldsymbol{a}^t\dfrac{W^0}{{\rm det}B}
\boldsymbol{a}\right).
\label{fizerofull}
\end{equation}
The wavefunction normalization is
\begin{equation}
\mathcal{A}=({\rm det}B)^{-1/2},
\label{det}
\end{equation}
and the time $\tau$ is defined by
\begin{equation}
\dfrac{d\tau}{d t}=\dfrac{1}{{\rm det}B}.
\label{tempo}
\end{equation}
The derivation of the scaling equations (\ref{sec:appA}) relies on the
particular choice of the above phase $\phi$ which verifies
\begin{equation}
  \boldsymbol{\nabla}_r\phi=-\dfrac{m}{\hbar}B\dfrac{\partial\boldsymbol{\rho}}{\partial
    t}\quad\text{or}\quad \boldsymbol{v}(\mathbf{r})=\dot{B}B^{-1}\,\mathbf{r} - \dot{B}B^{-1}\,\mathbf{r}^{\text{cm}}+B^{-1}\dot{\mathbf{r}}^{\text{cm}},
\end{equation}
$\boldsymbol{v}(\mathbf{r})$ being the velocity field of the
condensate, and on the assumption that the matrix $\dot B B^{-1}$ is
symmetric.  The first condition consists in imposing that there are no
terms linear in momentum in the GPE in the
$\boldsymbol{\rho}$-coordinate frame; if the first condition is
fulfilled the second imposes that the velocity field is irrotational,
namely that the condensate is a superfluid everywhere as well.  This
implies that our scaling ansatz does not take into account the
presence of quantized vortices and thus can describe the dynamics of a
rotated condensate only below the critical angular velocity $\dot\alpha_c\simeq 0.7\omega_x$ for a slightly
anisotropic confinement \cite{Madison2000}, or in general, for a
metastable configuration \cite{Recati2001}.  Nevertheless, a slightly
modified ansatz could be deviced to incorporate the possibility of
quantized vortices. It is also possible to relax the first condition
and allow for terms in the GPE that contain for instance the angular
momentum components. These extensions are deferred for future
studies.

Equations \eqref{scaling_full} and \eqref{shift_full} can be used to
determine the dipolar, compressional and scissors modes for a
harmonically-trapped superfluid condensate (see \ref{sec:appB}).  Replacing ${\rm det}B$
with $({\rm det}B)^\beta$ in Eq.  \eqref{scaling_full}, the same
equation describes the compression and the scissors dynamics of a
superfluid characterized by an equation of state $\mu(n)\propto
n^\beta$, as it has been already shown for the quadrupolar modes
\cite{Hui2004} and as it can be easily deduced by Eq.~\eqref{unpomenolunga} of the Appendix.  In the following we present
three possible shortcut trajectories based on these scaling equations
and adapted to compress or decompress and rotate a BEC in the absence and
in the presence of gravity.

\subsection{Uniform decompression or compression of a condensate}

We now consider the particular case of $\boldsymbol{u} = \mathbf{0}$
and $W$ diagonal. If one wants to compress or decompress the
condensate without modifying the condensate aspect ratio, the
condition $\omega_i(t_f)=\omega_i(0)/\gamma^2$ must hold for any
$i$. The boundary conditions for the shortcut solution are: $\dot
b_{ii}(0)=\dot b_{ii}(t_f)=0$, $b_{ii}(0) = 1$,
$b_{ii}(t_f)=\gamma^{4/5}$ and $\ddot b_{ii}(0)=\ddot b_{ii}(t_f)=0$.
One possible solution is to set all $b_{ii}(t)$'s equal to a unique
function
\begin{equation}
b(t)=\sum_{k=0}^{5}c_k\left(\frac{t}{t_f}\right)^k
\end{equation}
with $c_0=1$, $c_1=c_2=0$, $c_3=10(\gamma^{4/5}-1)$, $c_4=-15(\gamma^{4/5}-1)$, $c_5=6(\gamma^{4/5}-1)$.  The time evolution of the trap frequencies $\omega_i(t)$ will be given by the equation
\begin{equation}
\omega_i^2(t)=\frac{\omega_i^2(0)}{b^5}-\frac{\ddot b}{b}.
\end{equation}
If the kinetic energy is negligible during the whole decompression,
the final state is a BEC at equilibrium with a chemical potential that
has been divided by a factor of $\gamma^{16/5}$ (because $\mu \propto
\left(\Pi_i \omega_i \right)^{2/5}$).

\subsection{General compression or decompression in the presence of gravity}

We now consider the case where $W(t)$ is diagonal with
$\omega_x(t)=\omega_z(t)=\omega_\perp(t)$,
$\omega_y(t)=\omega_\parallel(t)$, and $u_z = mg$. A general
compression or decompression of a condensate confined in this
axially-symmetric trap \eqref{trap} can be realized in two steps: (i)
in the first step ($t \in [0, \bar{t}\,]$), $b_\parallel$ is kept
fixed as outlined in Sec.~\ref{sec:interacting_bec}, while the desired
final value of $b_{\perp} = b_{\perp}(t_f)$ ($R_{\perp}(t_f)$) is
reached; (ii) then ($t \in [\bar{t}, t_f]$) $b_\perp$ is fixed and
$b_\parallel$ evolves according to the set of equations:
\begin{gather}
\omega_\perp^2(t)=\dfrac{\omega_{\perp}^2(\bar{t})}{b_\parallel(t)},\\
\ddot b_\parallel(t)+b_\parallel(t)\omega_\parallel^2(t)=
\dfrac{\omega_{\parallel}^2(\bar{t})}{b_\parallel^2(t)},\\
b_\parallel(t)\ddot c(t)=\omega_{\perp}^2(\bar{t})\left(c(t)-b_\parallel(t)\right),
\label{cm_bpar_fix}
\end{gather}
where $c(t) = -\omega_{\perp}^2(\bar{t}) r_z^\text{cm}(t)/(g b_\perp(t))$
as in Eq. \eqref{ciditi}.  Also in this case one can write the
function $c(t)$ as a polynomial of order $\ge 9$ (see Eq.
\eqref{eq:ansatz}) with the first coefficient fixed to one and the
following four coefficients fixed to zero. The other coefficients are
fixed by the boundary conditions at the time $t_f$ of the function
$c(t)$ and of the function $b_\parallel(t)$, that from Eq.
\eqref{cm_bpar_fix} can be written as
\begin{equation}
b_\parallel(t)=-\dfrac{\omega_{\perp}^2(\bar{t}) c(t)}{\ddot c(t)-\omega_{\perp}^2(\bar{t})},
\end{equation}
and by the boundary conditions of their derivatives at the same time $t_f$.

\subsection{Rotation of the BEC in the presence of gravity}

Now we propose a shortcut trajectory to rotate an axially-symmetric
BEC of an angle $\bar\alpha$, in the presence of the gravity.  In this
case
\begin{equation}
 W(0)=\left(\begin{tabular}{ccc}
$\omega_\perp^2(0)$ &0&0\\
0&$\omega_{\parallel}^2(0)$&0\\
0&0&$\omega_{\perp}^2(0)$\\
\end{tabular}
\right),
\end{equation}
and $W(t_f)=R^{-1}_{\bar\alpha}W(0)R_{\bar\alpha}$, with
\begin{equation}
R_{\bar\alpha}=\left(\begin{array}{ccc}
1&0&0\\
0&\cos\bar\alpha &\sin\bar\alpha\\
0&-\sin\bar\alpha&\cos\bar\alpha
\end{array}
\right).
\end{equation}
Let us suppose, for instance, $\omega_{\perp}(0)<\omega_{\parallel}(0)$, with
$\omega_{\parallel}(0) = \lambda\omega_{\perp}(0)$. The tilted ground-state
for the potential $W(t_f)$ can be obtained in two steps: (i) during a
time $\bar t$, fixing $b_\parallel$, decompressing the BEC in the
radial direction up to the value $b_\perp(\bar t)=\lambda^{-1}$. At
$t=\bar t$ the trap is spherical with frequency
$\tilde\omega=\lambda\omega_{\parallel}(0)$ and the BEC is spherical
with a TF radius equals to $R_\parallel(0)$.  (ii) Fixing
$b_\parallel$ along the direction $y'$, compressing in the direction
$x'$ and $z'$, where the axis $\mathbf{r}'$ are defined by
$\mathbf{r}' = R_{\bar\alpha}\mathbf{r}$.  Using the new
coordinate reference frame, and setting $c_z(t) =
-{\tilde\omega}^2r_z^\text{cm}(t)/(gb_\perp(t) \cos{\bar\alpha})$, and
$c_y(t)= -{\tilde\omega}^2r_y^\text{cm}(t)/(g \sin{\bar\alpha})$, we
obtain the set of equations
\begin{gather}
\label{eq:bperp1_rot}
\ddot b_\perp(t)+ b_\perp(t)\omega_\perp^2(t) = 
\tilde\omega^2/b_\perp^3(t) , \\
\label{eq:bparallel1_rot}
\omega_\parallel(t) = 
\tilde\omega/b_\perp(t) , \\
\label{eq:a1_rot}
b_\perp^4(t)\ddot c_z(t)+2b_\perp^3(t)
\dot b_\perp(t)\dot c_z(t)
+{\tilde\omega}^2(c_z(t)-b_\perp^3(t))=0 , \\
b_\perp^2(t)\ddot c_y(t)+{\tilde\omega}^2 (c_y(t)-b_\perp^2(t))=0 ,
\end{gather}
the latter describing the centre-of-mass motion in the $y'$ direction.
The boundary conditions for such a problem are: $b_\perp(\bar
t)=c_z(\bar t)=c_y(\bar t)=1$, $b_\perp(t_f)=\lambda$,
$c_z(t_f)=\lambda^3$, $c_y(t_f)=\lambda^2$, and that all the first and
the second derivatives with respect to time are null at $t=\bar t$ and
$t_f$. In this case a finite-order polynomial ansatz in
$\tau$ for $c_i$ was found to be inadequate as a solution of the
scaling equations due to the coupling of $c_y$ and $c_z$. A full
numerical solution of the dynamical equation using, e.g., a shooting
method \cite{NR} or following a strategy as that implemented in
optimal control \cite{Hohenester2007} may be needed in finding a
shortcut trajectory in this case.

\section{Conclusion}

We have experimentally demonstrated the controlled transfer of trapped
ultracold atoms between two stationary states using a
faster-than-adiabatic process which reduces the transfer time down to
a few tens of milliseconds. The transfer is achieved by engineering
specific trajectories of the external trapping frequencies that take
explicitly into account the spatial shift introduced by gravity.  This
scheme was successfully applied both to a thermal gas of atoms and to
an almost pure Bose-Einstein condensate. The scheme used is flexible
enough to be adapted to both situations even though in the thermal gas
interactions does not play a significant role while the Bose-Einstein
condensate is strongly affected by the $s$-wave scattering of atoms.  
The residual excitations observed after the shortcut decompressions in the present demonstration experiments are due to our imperfect control over the time-varying magnetic trapping potential, and could be substantially reduced in future realizations.

Theoretically, the design of the transfer process was based on the
invariants of motion and scaling equations techniques which turned out
to be possible thanks to the harmonic shape of the external potential.
In our scheme,
the invariant of motion technique (for non-interacting particles) and
the scaling equations technique (valid for both the non-interacting
and the interacting gas) are tightly connected. The invariant of
motion we used is a time-independent harmonic oscillator Hamiltonian
that can be obtained by a time-dependent canonical transformation of
position and momentum.  In the scaling equation technique, we looked
for a scaling plus shift transformation of the coordinate that
allowed the equation of motion for the system to be time-independent
(except for terms that are not coordinate-dependent).  In both cases
the whole dynamics is included in the new set of (canonical)
coordinates, that depend on the trap frequencies.  We also showed that
these techniques can be generalized to include the rotation of the
eigenaxes without much effort.

Very often, in cold-atom experiments, samples are prepared by
transferring atoms from some confinement to another, e.g., from a
magneto-optical trap to a magnetic quadrupolar trap, from a
quadrupolar trap to a Ioffe-Pritchard trap, from an harmonic
confinement to an optical lattice, etc., the main limitation being
that, for short transfer times, parasitic excitations may show up. The
main application of our scheme is to guide this transfer in order to
prepare a very cold sample in a very short time with the desired
geometry and without exciting unwanted modes.  The
shortcut-to-adiabacity scheme proposed here could be applied to
non-interacting particles such as cold gases or ultracold
spin-polarized fermions, to normal or superfluid (bosonic or fermionic
as well) gases in the hydrodynamic regimes, and to strongly correlated
systems such as the Tonks gas.  In this paper we focused on explicit
solutions to transfer atoms between two stable states, but the same
strategy could be applied to control the generation of metastable
states, vortex states, or some exotic out-of-equilibrium states.  We
plan to explore these possibilities in future studies.

\acknowledgments{This work was supported by CNRS and Universit\'e de
  Nice-Sophia Antipolis. We also acknowledge financial support from
  R\'egion PACA, F\'ed\'eration Wolfgang Doeblin, and CNRS-CONICET
  international cooperation grant n$^\circ$22966. JFS acknowledges
  support from the French ministry of research and education for his
  funding, and thanks Mario Gattobigio and Michel Le~Bellac for
  helpful discussions on theoretical aspects.}

\appendix

\section{Demonstation of Eqs.~\eqref{scaling_full} and \eqref{shift_full}}
\label{sec:appA}

In this appendix we derive Eqs. \eqref{scaling_full}-\eqref{tempo}.  The
starting point is the GPE \eqref{eq:gpe} for a general potential
\eqref{generalpot}. We look for a solution of the form
\begin{equation}
\psi(\mathbf{r},t)=\mathcal{A}(t) \chi(\boldsymbol{\rho},\tau)e^{i\phi(\mathbf{r},t)}
\end{equation}
with
\begin{equation}
\boldsymbol{\rho}=B^{-1}\mathbf{r}+\boldsymbol{a}.
\end{equation}
Equation \eqref{eq:gpe} then takes the form
\begin{equation}\begin{split}
    i\hbar
    \left[\dfrac{\dot{\mathcal{A}}}{\mathcal{A}}\chi+\nabla_{\boldsymbol{\rho}}\chi\cdot\dfrac{\partial\boldsymbol{\rho}(B,\boldsymbol{a})}{\partial
        t}+
      \dfrac{\partial\chi}{\partial \tau}\dfrac{\partial\tau}{\partial t}+i\chi\dot\phi \right]=\\
    -\dfrac{\hbar^2}{2m}\left\{\sum_{i,j,k}[B^{-1}]_{ij}[B^{-1}]_{kj}\dfrac{\partial^2\chi}{\partial\rho_i\partial\rho_k}+2i(B^{-1}\nabla_{\mathbf{r}}\phi)\cdot\nabla_{\boldsymbol{\rho}}\chi+i(\nabla^2_{\mathbf{r}}\phi)\chi-\left(\nabla_{\mathbf{r}}\right)^2\chi\right\}\\
    +\dfrac{1}{2}m\left\{ [B(\boldsymbol{\rho}-\boldsymbol{a})]^tW
      [B(\boldsymbol{\rho}-\boldsymbol{a})]\right\}\chi+\boldsymbol{u}^tB(\boldsymbol{\rho}-\boldsymbol{a})\chi+g|\mathcal{A}|^2|\chi|^2\chi.
\end{split}
\label{lllongeq}
\end{equation}
We look for the conditions that $\mathcal A,B$, and $\boldsymbol{a}$
have to verify aiming to simplify Eq. \eqref{lllongeq} to the form
\begin{equation}
i \hbar \frac{\partial}{\partial \tau} \chi(\boldsymbol{\rho}, \tau) = 
\left[ U(\boldsymbol{\rho},0) + 
\tilde{V} N |\chi(\boldsymbol{\rho}, \tau)|^2 \right]\chi(\boldsymbol{\rho}, \tau), 
\end{equation}
in the TF limit, namely, neglecting the kinetic term given in Eq.~\eqref{brutto}. We deduce immediately that (i) the second term of Eq.
\eqref{lllongeq} has to be equal to the sixth, and (ii) the first to
the seventh.  Condition (i) leads to
\begin{equation}
\nabla_{\mathbf{r}}\phi=-\dfrac{m}{\hbar}B\left\{\dot{[B^{-1}]}\mathbf{r}
+\dot{\boldsymbol{a}}\right\},
\end{equation}
that has a solution if the matrix $B\dot{[B^{-1}]}=-\dot{B}B^{-1}$ is
symmetric\footnote{In a general case the matrix $\dot BB^{-1}$ can be split into a
symmetric and an antisymmetric part. In the
$\boldsymbol{\rho}$-frame of reference, the antisymmetric part gives rise to a
rotational term proportional to the angular momentum and
only the symmetric part of $\dot BB^{-1}$ contributes to the phase
of the wave function.   The rotational term can be neglected for
nearly-isotropic trap or for small angular velocities of the trap.}. 
If this condition holds, we get Eq. \eqref{phase_gen} for the phase $\phi$.
Condition (ii) can be written as
\begin{equation}
\dot{\mathcal{A}}{\mathcal{A}}^{-1}=-\dfrac{1}{2}{\rm tr}(\dot BB^{-1}).
\end{equation}
Using the invariance of the trace and of the determinant, the
evolution of $\mathcal{A}$ can be rewritten in term of the eigenvalues
$\beta_i$' of the matrix $B$ as
\begin{equation}
\begin{split}
\dot{\mathcal{A}}{\mathcal{A}}^{-1}=-\dfrac{1}{2}\sum_i\dfrac{\dot\beta_i}{\beta_i}=-\dfrac{1}{2}\dfrac{\partial}{\partial t}\ln{\rm det} B\\
\dfrac{\partial}{\partial t} \ln \mathcal{A}=-\dfrac{1}{2}\dfrac{\partial}{\partial t}\ln{\rm det} B
\end{split}
\label{eq:det0}
\end{equation}
If, e.g., at $t=0$ we have that $B$  is the identity and
$\mathcal{A}=1$, equation~\eqref{eq:det0} yields Eq.~\eqref{det}.

Moreover from the comparison between the third term in Eq.
\eqref{lllongeq} and the non-linear term (condition (iii)), we deduce
Eq. \eqref{tempo}.  Taking into account (i)-(iii), Eq.
\eqref{lllongeq} reduces to
\begin{equation}
\begin{split}
&i\hbar \dfrac{\partial\chi}{\partial \tau}-
\hbar{\rm det}B\dfrac{\partial\phi_0}{\partial t}=\\
&{\rm det}B\left\{\dfrac{m}{2}\left[\dot{B}B^{-1}\mathbf{r}-B\dot{\boldsymbol{a}}^2 +\mathbf{r}^t\ddot{B}B^{-1}\mathbf{r}+
\mathbf{r}^t\dot{B}\dot{[B^{-1}]}\mathbf{r}\right]\right.\\
&\left.-m\mathbf{r}^t\dot B\dot{\boldsymbol{a}}-
m\mathbf{r}^tB\ddot{\boldsymbol{a}}+
\dfrac{1}{2}m\left\{ [B(\boldsymbol{\rho}-\boldsymbol{a})]^tW
[B(\boldsymbol{\rho}-\boldsymbol{a})]\right\}+
\boldsymbol{u}^tB(\boldsymbol{\rho}-\boldsymbol{a})\right\}\chi\\
&+g|\chi|^2\chi.
\end{split}
\label{unpomenolunga}
\end{equation}
By imposing the quadratic term in $\boldsymbol{\rho}$ to be equal to
$\frac{m}{2}{\boldsymbol{\rho}}^tW^0\boldsymbol{\rho}$, we get
condition (iv), i.e., Eq.  \eqref{scaling_full}; the fifth condition
is that the linear term in $\boldsymbol{\rho}$ vanishes and thus leads
to \eqref{shift_full}; finally by requiring that the
$\boldsymbol{\rho}$-independent term be null, we get
\eqref{fizerofull} for $\phi_0$.

\section{Low-lying modes}
\label{sec:appB}

Equation~\eqref{shift_full} describes the dipolar modes for the centre of mass
and Eq.~\eqref{scaling_full} the quadrupolar and the scissors modes.
The low-lying eigenfrequencies of these latter modes can be obtained by
solving the equation of motion for the matrix $B$ for the case of a tilt
of the trap of a small angle $\alpha$. At $t>0$, the matrix $W$ is constant
and can be written as
\begin{equation}
W=\left(\begin{tabular}{ccc}
$\omega_\perp^2$ &0&0\\
0&$\omega_\parallel^2$&$\alpha(\omega_\parallel^2-\omega_\perp^2)$\\
0&$\alpha(\omega_\parallel^2-\omega_\perp^2)$&$\omega_\perp^2$\\
\end{tabular}
\right)=W^0+\delta W
\end{equation}
where
\begin{equation}
W^0=\left(\begin{tabular}{ccc}
$\omega_\perp^2$ &0&0\\
0&$\omega_\parallel^2$&0\\
0&0&$\omega_\perp^2$\\
\end{tabular}
\right)
\end{equation}
and 
\begin{equation}
\delta W=\left(\begin{tabular}{ccc}
0 &0&0\\
0&0&$\alpha(\omega_\parallel^2-\omega_\perp^2)$\\
0&$\alpha(\omega_\parallel^2-\omega_\perp^2)$&0\\
\end{tabular}
\right).
\end{equation}
We look for solutions of the form $B^t=1+\delta$.
Equation \eqref{scaling_full} takes the form:
\begin{equation}
\ddot\delta\simeq -W^0\delta-\delta^t W^0-({\rm Tr}\delta) W^0+ \delta W,
\label{firstorder}
\end{equation}
at the first order in $\delta$.
For the diagonal terms we have
\begin{equation}
\ddot\delta_{ii}=-2\omega_i^2\delta_{ii}-({\rm Tr}\delta) \omega_i^2.
\end{equation}
Setting $\delta_{ii}=\Delta_ie^{i\Omega t}$, we obtain the following
coupled equations
\begin{equation}
\begin{split}
-\Omega^2 \Delta_x=-2\omega_\perp^2\Delta_x-(\Delta_x+\Delta_y+\Delta_z)\omega_\perp^2\\
-\Omega^2 \Delta_y=-2\omega_\parallel^2\Delta_y-(\Delta_x+\Delta_y+\Delta_z)\omega_\parallel^2\\
-\Omega^2 \Delta_z=-2\omega_\perp^2\Delta_z-(\Delta_x+\Delta_y+\Delta_z)\omega_\perp^2\\
\end{split}
\end{equation}
whose solutions are the surface mode
$\Omega=\sqrt{2}\omega_\perp$ for any values of $\omega_\perp$ and 
$\omega_\parallel$, and  the breathing modes
$\Omega\simeq 2\omega_\perp$ and $\Omega\simeq \sqrt{5/2}\omega_\parallel$ in the cigar-shape regime $\omega_\parallel\ll\omega_\perp$.

For the off-diagonal terms $\delta_{ij}$, ($\{i,j\}=\{2,3\}$ or $\{3,2\}$),
Eq. \eqref{firstorder} gives
\begin{equation}
\ddot\delta_{ij}=-\omega_i^2\delta_{ij}-\omega_j^2\delta_{ji}+\alpha(\omega_\parallel^2-\omega_\perp^2)
\end{equation}
namely
\begin{equation}
\ddot\delta_{ij}+\ddot\delta_{ji}=-(\omega_i^2+\omega_j^2)(\delta_{ij}+\delta_{ji})+2\alpha(\omega_\parallel^2-\omega_\perp^2).
\label{off}
\end{equation}
Equation \eqref{off} has solution 
\begin{equation}
\delta_{23}=\delta_{32}=\alpha\dfrac{(\omega_\parallel^2-\omega_\perp^2)}{\Omega_s^2}[1-\cos(\Omega t)],
\end{equation}
where $\Omega_s=(\omega_\perp^2+\omega_\parallel^2)^{1/2}$. This is a
scissors mode with boundary conditions $\dot \delta_{ij}(t=0)=0$ and
$\delta_{ij}(t=0)=0$.

\bibliographystyle{unsrtnat}  
\bibliography{biblio}

\end{document}